\documentclass[12pt]{article}
\renewcommand{\theequation}{2.\arabic{equation}}
\begin{document}
\begin{center}
{\large \bf Light Cone Current Algebra\footnote[1]{Work supported in part
by the U.S. Atomic Energy Commission under contract AT(11--1)--68,
San Francisco Operations Office.}}
\end{center}
\bigskip
\begin{center}

{\bf Harald Fritzsch\footnote[7]{Max--Planck--Institut f\"ur Physik und
Astrophysik, M\"unchen, Germany. Present address (1971--1972): CERN,
Geneva, Switzerland.}}
\end{center}
\begin{center}
{\bf and}
\end{center}
\begin{center}
{\bf Murray Gell--Mann}\footnote[2]{California Institute of Technology, Pasadena,
California. Present address: (1971--1972): CERN, Geneva, Switzerland.}
\end{center}
\bigskip

\begin{abstract} This talk follows by a few months a talk by the same authors on
nearly the same subject at the Coral Gables Conference. The ideas
presented here are basically the same, but with some amplification, some
change of viewpoint, and a number of new questions for the future. For our
own convenience, we have transcribed the Coral Gables paper, but with an
added ninth section, entitled ``Problems of light cone current algebra'',
dealing with our present views and emphasizing research topics that
require study.
\end{abstract}
\bigskip
\bigskip
{\bf 1. INTRODUCTION}\\
{\small We should like to show that a number of different ideas of the
last few years on broken scale invariance, scaling in deep inelastic
electron--nucleon scattering, operator product expansions on the light
cone, ``parton'' models, and generalizations of current algebra, as well
as some new ideas, form a coherent picture. One can fit together the
parts of each approach that make sense and obtain a consistent view of
scale invariance, broken by certain terms in the energy density, but
restored in operator commutators on the light cone.\\
We begin in the next section with a review of the properties of the
dilation operator $D$ obtained from the stress--energy--momentum tensor
$\Theta_{\mu \nu}$ and the behavior of operators under equal--time
commutation with $D$, which is described in terms of physical dimensions
$l$ for the operators. We review the evidence on the relation between the
violation of scale invariance and the violation of $SU_3 \times SU_3$
invariance.\\
Next, in Section 3, we describe something that may seem at first sight
quite different, namely the Bjorken scaling of deep inelastic scattering
cross sections of electrons on nucleons and the interpretation of this
scaling in terms of the light cone commutator of two electromagnetic
current operators. We use a generalization of Wilsons's work$^1$, the
light--cone expansion emphasized particularly by Brandt and Preparata$^2$
and Frishman$^3$. A different definition $l$ of physical dimension is
thus introduced and the scaling implies a kind of conservation of $l$ on
the light cone. On the right--hand side of the expansions, the operators
have $l = -J - 2$, where $J$ is the leading angular momentum contained in
each operator and $l$ is the leading dimension.\\
In Section 4, we show that under simple assumptions the dimensions $l$ and
$\bar{l}$ are essentially the same, and that the notions of scaling and
conservation of dimension can be widely generalized. The essential
assumption of the whole approach is seen to be that the dimension $l$ or
($\bar{l}$) of any symmetry--breaking term in in the energy (whether
violating scale invariance or $SU_3 \times SU_3$) is {\it higher} than
the dimension, $-4$, of the completely invariant part of the energy
density. The conservation of dimension on the light cone then assigns a
lower singularity to symmetry--breaking terms than to symmetry--preserving
terms, permitting the light--cone relations to be completely symmetrical
under scale, $SU_3 \times SU_3$, and perhaps other symmetries.\\
\\
In Section 5, the power series expansion on the light cone is formally
summed to give bilocal operators (as briefly discussed by Frishman) and it
is suggested that these bilocal light--cone operators may be very few in
number and may form some very simple closed algebraic system. They are
then the basic mathematical entities of the scheme.\\
It is pointed out that several features of the Stanford experiments, as
interpreted according to the ideas of scaling, resemble the behavior on the
light cone of free field theory or of interacting field theory with naive
manipulations of operators, rather than the behavior of renormalized
perturbation expansions of renormalizable field theories. Thus free field
theory models may be studied for the purpose of abstracting algebraic
relations that might be true on the light cone in the real world of
hadrons. (Of course, matrix elements of operators in the real world
would not in general resemble matrix elements in free field theory.) Thus
in Section 6 we study the light--cone behavior of local and bilocal
operators in free quark theory, the simplest interesting case. The
relavant bilocal operators turn out to be extremely simple, namely just
$i/2\left( \bar{q}(x)\lambda_i \gamma_{\alpha} q (y) \right)$ and
$i/2 \left( \bar{q} (x) \lambda_i \gamma_{\alpha} \gamma_5 q (y) \right)$,
bilocal generalizations of $V$ and $A$ currents. The algebraic system to
which they belong is also very simple.\\
In Section 7 we explore briefly what it would mean if these algebraic
relations of free quark theory were really true on the light cone for
hadrons. We see that we obtain, among other things, the sensible features
of the so--called ``parton'' picture of Feynman$^4$ and of Bjorken and
Paschos$^5$, especially as formulated more exactly by Landshoff and
Polkinghorne$^6$, Llewellyn Smith$^7$, and others. Many symmetry relations
are true in such a theory, and can be checked by deep inelastic experiments
with electrons and with neutrinos. Of course, some alleged results of
the ``parton'' model depend not just on light cone commutators but on
detailed additional assumptions about matrix elements, and about such
results we have nothing to say.\\
The abstraction of free quark light cone commutation relations becomes
more credible if we can show, as was done for equal time charge density
commutation relations, that certain kinds of non--trivial interactions
of quarks leave the relations undisturbed, according to the method of
naive manipulation of operators, using equations of motion. There is
evidence that in fact this is so, in a theory with a neutral scalar or
pseudoscalar ``gluon'' having a Yukawa interaction with the quarks.
(If the ``gluon'' is a vector boson, the commutation relations on the
light cone might be disturbed for all we know.)\\
A special case is one in which we abstract from a model in which there
are only quarks, with some unspecified self--interaction, and no
``gluons''. This corresponds to the pure quark case of the ``parton''
model. One additional constraint is added, namely the identification of
the traceless part of $\Theta_{\mu\nu}$ with the analog of the traceless
part of the symmetrized $\bar{q} \gamma_{\mu} \partial_{\nu} q$. This
constraint leads to an additional sum rule for deep inelastic electron
and neutrino experiments, a rule that provides a real test of the pure
quark case.\\
We do not, in this paper, study the connection between scaling in
electromagnetic and neutrino experiments on hadrons on the one hand and
scaling in ``inclusive'' reactions of hadrons alone on the other hand.
Some approaches, such as the intuition of the ``parton''  theorists,
suggest such a connection, but we do not explore that idea here. It is
worth reemphasizing, however, that any theory of pure hadron behavior
that limits transverse momenta of particles produced at high energies
has a chance of giving the Bjorken scaling when electromagnetism and
weak interactions are introduced. (This point has been made in the
cut--off models of Drell, Levy, and Yan$^8$).\\
\\
\\
{\bf 2. DILATION OPERATOR AND BROKEN SCALE INVARIANCE$^9$}\\
\\
We assume that gravity theory (in first order perturbation approximation)
applies to hadrons on a microscopic scale, although no way of checking
that assertion is known. There is then a symmetrical, conserved, local
stress--energy--momentum tensor $\Theta_{\mu \nu} (x)$ and in terms of it
the translation operators $P_{\mu}$, obeying for any operator
$O \ldots (x)$, the relation
\begin{equation}
\left[
O \ldots (x), P_{\mu} \right] = \frac{1}{i} \partial_{\mu} O \ldots (x),
\end{equation}
are given by
\begin{equation}
P_{\mu} = \int \Theta_{\mu 0} d^3 x.
\end{equation}
\\
Now we want to define a differential dilation operator $D(t)$ that
corresponds to our intuitive notions of such an operator, i. e., one that
on equal--time commutation with a local operator $O \ldots$ of definite
physical dimension $\bar{l}_0$, gives
\begin{equation}
\left[ O \ldots (x), D(t) \right] = i x_{\mu} \partial_{\mu} O \ldots
(x) - i \bar{l}_{\sigma} O \, \ldots (x) \, .
\end{equation}
\\
We suppose that gravity selects a $\Theta_{\mu \nu}$ such that this
dilation operation $D$ is given by the expression
\begin{equation}
D = - \int x_{\mu} \Theta_{\mu 0} d^3 x \, .
\end{equation}
\\
It is known that for any renormalizable theory this is possible, and
Callan, Coleman, and Jackiw have shown that in such a case the matrix
elements of this $\Theta_{\mu \nu}$ are finite. From (2.4) we see that
the violation of scale invariance is connected with the non--vanishing
of $\Theta_{\mu \nu}$ since we have
\begin{equation}
\frac{dD}{dt} = - \int \Theta_{\mu \mu} \, d^3 x \, .
\end{equation}
\\
Another version of the same formula says that
\begin{equation}
\left[ D, P_0 \right] = - i P_0 - i \int \, \Theta_{\mu \mu} \, d^3 x
\end{equation}
\\
and we see from this and (2.3) that the energy density has a main
scale--invariant term $ \stackrel{=}{\Theta}_{00}$ (under the complete
dilation operator $D)$ with $l = -4$ (corresponding to the mathematical
dimension of energy density) and other terms $w_n$ with other physical
dimensions $\bar{l}_n$. The simplest assumption (true of most simple
models) is that these terms are world scalars, in which case we obtain
\begin{equation}
- \Theta_{\mu \nu} = \sum\limits_{n} \left( \bar{l}_n + 4 \right) w_n
\end{equation}
along with the definition
\begin{equation}
\Theta_{00} = \stackrel{=}{\Theta}_{00} + \sum\limits_{n} w_n \, .
\end{equation}
\\
We note that the breaking of scale invariance prevents $D$ from being a
world scalar and that equal--time commutation with $D$ leads to a
non--covariant break--up of operators into pieces with different
dimensions $\bar{l}$.\\
To investigate the relation between the violations of scale invariance and
of chiral invariance, we make a still further simplifying assumption (true
of many simple models such as the quark--gluon Lagrangian model), namely
that there are two $q$--number $w$'s, the first violating scale invariance
but not chiral invariance (like the gluon mass) and the second violating
both (like the quark mass):
\begin{equation}
\Theta_{00} = \stackrel{=}{\Theta}_{00} + \delta + u + \, \, \,
{\rm const.} \, ,
\end{equation}
with $\delta $ transforming like $({\bf 1}, {\bf 1})$ under
$SU_3 \times SU_3$. Now how does $u$ transform? We shall start with the
usual theory that it all belongs to a single
$\left( {\bf 3}, \bar{{\bf 3}} \right) + \left( \bar {\bf 3}, {\bf 3}
\right)$ representation and that the
smallness of $m_{\pi}^2$ is to be attributed, in the spirit of {\bf PCAC},
to the small violation of $SU_2 \times SU_2$ invariance by $u$. In that
case we have
\begin{equation}
u = - u_0 - cu_8 \, ,
\end{equation}
with $c$ not far from $- \sqrt{2}$, the value that gives $SU_2 \times
SU_2$ invariance and $m_{\pi}^2 = 0$ and corresponds in a quark scheme to
giving a mass only to the $s$ quark. A small amount of $u_3$ may be
present also, if there is a violation of isotopic spin conservation that
is not directly electromagnetic; an expression containing $u_0, u_3$ and
$u_8$ is the most general canonical form of a $CP$--conserving term
violating $SU_3 \times SU_3$ invariance and transforming like
$ \left( {\bf 3}, {\bf \bar{3}} \right) + \left( {\bf \bar{3}},
{\bf 3} \right)$.\\
According to all these simple assumptions, we have
\begin{equation}
- \Theta_{\mu \nu} = \left( \bar{l}_{\delta} + 4 \right) \delta + \left(
\bar{l}_u + 4 \right) \left( - u_0 - cu_8 \right) + 4 \, \, \,
{\rm (const.)}
\end{equation}
and, since the expected value of $\left( - \Theta_{\mu \nu} \right)$ is
$2 m^2$, we have
\begin{equation}
0 = \left( \bar{l}_{\delta} + 4 \right) < {\rm vac} \, \mid \delta
\mid \, {\rm vac} \,  > + \left( \bar{l}_u + 4 \right) < \,
{\rm vac} \, \mid u \mid \, {\rm vac} \, > + 4 \, 
{\rm (const.)},
\end{equation}
\begin{eqnarray}
2 m^2_i \left( PS8 \right) & = & \left( l_{\delta} + 4 \right)
\left( PS_i \mid \delta \mid PS_i \right) \nonumber \\
\\
& & + \left( l_u + 4 \right) < PS_i \mid u \mid PS_i >, \nonumber
\end{eqnarray}
\\
etc.\\
The question has often been raised whether $\delta $ could vanish. Such a
theory is very interesting, in that the same term $u$ would break chiral
and conformal symmetry. But is it possible?\\
It was pointed out a year or two ago$^{10}$ that for this idea to work,
something would have to be wrong with the final result of von Hippel and
Kim$^{11}$, who calculated approximately the ``$\sigma $ terms'' in
meson--baryon scattering and found, using our theory of $SU_3 \times
SU_3$ violation, that $< N \mid U \mid N >$ was very small compared to
$2 m^2_N$. Given the variation of $< B \mid u_8 \mid B >$ over the
$1/2^+$ baryon octet, the ratio of $< \Xi \mid u \mid \Xi >$ to $< N \mid
u \mid N >$ would be huge if von Hippel and Kim were right, and this
disagrees with the value $m^2_{\Xi} / m_N^2$ that obtains if
$\delta = 0.$\\
Now, Ellis$^{12}$ has shown that in fact the method of von Hippel and Kim
should be modified and will produce different results, provided there is
a dilation. A dilation is a neutral scalar meson that dominates the
dispersion relations for matrix elements of $\Theta_{\mu \mu}$ at low
frequency, just as the pseudoscalar octet is supposed to dominate the
relations for $\partial_{\alpha} F^5_{i \alpha}$. We are dealing in the
case of the dilation, with PCDC (partially conserved dilation current)
along with PCAC (partially conserved axial vector current). If we have
PCAC, PCDC, and $\delta = 0$, we may crudely describe the situation by
saying that as $u \rightarrow 0$ we have chiral and scale invariance of
the energy, the masses of a pseudoscalar octet and a scalar singlet go to
zero, and the vacuum is not invariant under either chiral or scale
transformations (though it is probably $SU_3$ invariant). With the dilation,
we can have masses of other particles non--vanishing as $u \rightarrow 0$,
even though that limit is scale invariant.\\
Dashen and Cheng$^{13}$ have just finished a different calculation of the
``$\sigma $ terms'' not subject to modification by dilation effects, and
they find, using our description of the violation of chiral invariance,
that $< N \mid u \mid N >$ at rest is around $2 m^2_N$, a result perfectly
compatibe with the idea of vanishing $\delta $ and yielding in that case
a value $l_u \approx - 3$ (as in a naive quark picture, where $u$ is a
quark mass term!).\\
An argument was given last year $^{10}$ that if $\delta = 0$, the value of
$l_u$ would have to be $-2$ in order to preserve the perturbation theory
approach for $m^2 \left( PS 8 \right)$,
which gives the right mass formula for the pseudoscalar octet. Ellis,
Weisz, and Zumino$^{14}$ have shown that this argument can be evaded if
there is a dilation.\\
Thus at present there is nothing known against the idea that $\delta = 0$,
with $l_u$ probably equal to $-3$. However, there is no strong evidence in
favor of the idea either. Theories with non-vanishing $\delta $ operators
and various values of $l_{\delta}$ and $l_u$ are not excluded at all
(although even here a dilation would be useful to explain why
$< N \mid u \mid N >$ is so large). It is a challenge to theorists to
propose experimental means of checking whether the $\delta $ operator is
there or not.\\
It is also possible that the simple theory of chiral symmetry violation
may be wrong. First of all, the expression $- u_0 + \sqrt{2u_8}$ could be
right for the $SU_2 \times SU_2$--conserving but
$SU_3 \times SU_3$--violating part of $\Theta _{00}$, while the
$SU_2 \times SU_2$--violation could be accomplished by something quite
different from $\left( - c - \sqrt{2} \right) u_8$. Secondly, there can
easily be an admixture of the eighth component $g_8$ of an octet belonging
to (1,8) and (8,1). Thirdly, the whole idea of explaining $m_{\pi}^2
\approx 0$ by near--conservation of $SU_2 \times SU_2$ might fail, as
might the idea of octet violation of $SU_3$; it is those two hypotheses
that give the result that for $m_{\pi}^2 = 0$ we have only
$u_0 - \sqrt{2}u_8$ with a possible admixture of $g_8$. Here again there
is a challenge to theoreticians to propose effective experimental tests of
the theory of chiral symmetry violation.\\
\\
\\
{\bf 3. LIGHT CONE COMMUTATIONS AND DEEP INELASTIC}\\
{\bf ELECTRON SCATTERING}\\
\\
We want ultimately to connect the above discussion of physical dimensions
and broken scale invariance with the scaling described in connection with
the Stanford experiments on deep inelastic electron scattering$^{15}$. We
must begin by presenting the Stanford scaling in suitable form. For the
purpose of doing so, we shall assume for convenience that the experiments
support certain popular conclusions, even though uncertainties really
prevent us from saying more than that the experiments are consistent with
such conclusions:
\begin{enumerate}
\item[1)]that the scaling formula of Bjorken is really correct, with no
logarithmic factors, as the energy and virtual photon mass go to infinity
with fixed ratio;
\item[2)] that in this limit the neutron and proton behave differently;
\item[3)] that in the limit the longitudinal cross section for virtual
photons goes to zero compared to the transverse cross section.
\end{enumerate}
All these conclusions are easy to accept if we draw our intuition from
certain field theories without interactions or from certain field theories
with naive manipulation of operators. However, detailed calculations
using the renormalized perturbation expansion in renormalizable field
theories do not reveal any of these forms of behavior, unless of course
the sum of all orders of perturbation theory somehow restores the simple
situation. If we accept the conclusions, therefore, we should probably not
think in terms of the renormalized perturbation expansion, but rather
conclude, so to speak that Nature reads books on free field theory, as
far as the Bjorken limit is concerned.\\
To discuss the Stanford results, we employ a more or less conventional
notation. The structure functions of the nucleon are defined by matrix
elements averaged over nucleon spin,
\setcounter{page}{1}
\setcounter{equation}{0}
\renewcommand{\theequation}{3.\arabic{equation}}
\begin{eqnarray}
& & \frac{1}{4 \pi} \int d^4 x < N, p \mid \left[ j_{\mu} (x), j_{\nu}
(y) \right] \mid N, p > e^{-iq \cdot (x - y)} \nonumber \\
\nonumber \\
& = & \left( \delta_{\mu \nu} - \frac{q_{\mu} q_{\nu}}{q^2} \right)
W_1 \left( q^2, p \cdot q \right) \nonumber \\
\nonumber \\
& & + \left( p_{\mu}- \frac{p \cdot q}{p^2} q_{\mu} \right)
\left( p_{\nu} - \frac{p \cdot q}{q^2} q_{\nu} \right) W_2
\left( q^2, p \cdot q \right) \\
& = & \left( \delta_{\mu \nu} - \frac{q_{\mu} q_{\nu}}{q^2} \right)
\left( W_1 - \frac{ \left( p \cdot q \right)^2}{q^2} W_2 +
\frac{\delta_{\mu \nu} ( p \cdot q)^2 + p_{\mu} p_{\nu} q^2
- \left( p_{\mu \nu} + q_{\mu} q_{\nu} \right) p \cdot q}{q^2} W_2 \right)
\nonumber
\end{eqnarray}
\\
where $p$ is the nucleon four--momentum and $q$ the four--momentum of the
virtual photon. As $q^2$ and $q \cdot p$ become infinite with fixed ratio,
averaging over the nucleon spin and assuming $\sigma_L / \sigma_T
\rightarrow 0$, we can write the Bjorken scaling in the form
\begin{eqnarray}
\frac{1}{4 \pi} \int d^4 x & < & N, p \mid \left[ j_{\mu} (x),
j_{\nu}(y) \right] \mid N, p > e^{-i q \cdot (x - y)} \nonumber \\
\nonumber \\
& \rightarrow & \frac{\left( p_{\mu} p_{\nu} + p_{\nu} q_{\mu} \right)
p \cdot q - \delta_{\mu \nu} (p \cdot q)^2 -
p_{\mu} p_{\nu} q^2}{q^2 (q \cdot p)}{F_2 (\xi)},
\end{eqnarray}
where $\xi = - q^2 / 2 p \cdot q$ and $F_2 (\xi)$ is the scaling function
in the deep inelastic region.\\
\\
In coordinate space, this limit is achieved by approaching the light cone\\
$(x - y)^2 = 0^{15}$, and we employ a method, used by Frishman$^3$
 and by Brandt and Preparata,$^2$, generalizing earlier work of Wilson,
 that starts with an expansion for commutators or operator products valid
 near $(x - y)^2 = 0$. (The symbol $\hat{=}$ will be employed for equality
 in the vicinity of the light cone.) After the expansion is made, then the
 matrix element is taken between nucleons. To simplify matters, let us
 introduce the ``barred product'' of two operators, which means that we
 average over the mean position $R \equiv (x + y)/2$, leaving a function
 of $z \equiv x - y$ only (as appropriate for matrix elements with no
 change of momentum) and that we retain in the expansion only totally
 symmetric Lorentz tensor operators (as appropriate for matrix elements
 averaged over spin). Then the assumed light--cone expansion of the
 barred commutator $\overline{\left[ j_{\mu} (x), j_{\nu} (y) \right]}$
 tells us that we have, as $z^2 \rightarrow 0$,
 \begin{eqnarray}
 \overline{ \left[ j_{\mu} (x), j_{\nu} (y) \right]}
 & \hat{=} &
 t_{\mu \nu \rho \sigma} \left\{ \varepsilon \left( z_0 \right) \delta
 \left( z^2 \right) \left( O_{\rho \sigma} + \frac{1}{2!} z_{\alpha}
 z_{\beta} O_{\rho \sigma \alpha \beta} + \cdots \right)  \right\}
 \nonumber \\   \\
 & & + \left( \partial_{\mu} \partial_{\nu} - \delta_{\mu \nu}
 \partial^2 \right) \left\{ \varepsilon \left( z_0 \right) \delta
 \left( z^2 \right) \left( U + \frac{1}{2!} z_{\alpha} z_{\beta}
 U_{\alpha \beta} + \cdots \right) \right\}, \nonumber
 \end{eqnarray}
 where\\
 \begin{displaymath}
 t_{\mu \nu \rho \sigma} =
 \end{displaymath}
 \begin{displaymath}
 \frac{1}{\pi i} \left( \frac{2 \delta_{\mu \nu}\partial_{\mu}
 \partial_{\sigma} - \delta_{\rho \mu} \partial_{\nu} \partial_{\sigma}
 - \delta_{\rho \nu} \partial_{\mu} \partial_{\sigma} - \delta_{\sigma \mu}
 \partial_{\nu} \partial_{\rho} - \delta_{\sigma \mu} \partial_{\mu}
 \partial_{\sigma} - \delta_{\mu \sigma} \delta_{\nu \rho} \partial^2 -
 \partial_{\nu \sigma} \delta_{\mu \rho} \partial^2}{ \partial^2} \right)
 \end{displaymath}
 \\
 and the second term, the one that gives $\sigma _L$. will be ignored for
 simplicity in our further work.\\
 In order to obtain the Bjorken limit, we have only to examine the matrix
 elements between $\mid N p >$ and itself of the operators
 $0_{\alpha \beta}, 0_{\alpha \beta \gamma \delta},
 0_{\alpha \beta \gamma \delta} \varepsilon_{\rho},$ etc. The leading
 tensors in the matrix elements have the form $c_2p_{\alpha} p_{\beta},
 c_4 p_{\alpha} p_{\beta} p_{\gamma} p_{\delta}$ etc., where the c's
 are dimensionless constants. The lower tensors, such as
 $\delta _{\alpha \beta}$, have coefficients that are positive powers of
 masses, and these tensors give negligible contributions in the Bjorken
 limit. All we need is the very weak assumption that $c_2, c_4, c_8$, etc.,
 are not all zero, and we obtain the Bjorken limit.\\
 \\
 We define the function
 \begin{equation}
 \tilde{F} ( p \cdot z) = c_2 + \frac{1}{2!} \cdot c_4 (p \cdot c)^2 +
 \cdots \, .
 \end{equation}
 \\
 Taking the Fourier transform of the matrix elements of (3.3), we get
 in the Bjorken limit
 \begin{eqnarray}
 W_2 & \rightarrow & \frac{1}{2 \pi^2 i} \int d^4 ze^{-iq \cdot z}
 \tilde{F} (p \cdot z) \varepsilon \left( z_0 \right) \delta
 \left( z^2 \right) \nonumber \\ \nonumber \\
 & = & \frac{1}{2 \pi^2 i} \int F (\xi) d \xi \int d^4 z
 e^{-i\left(q+ \xi p \right) \cdot z} \varepsilon
 \left( z_0 \right) \delta \left( z^2 \right) \nonumber \\   \\
 & = & 2 \int F ( \xi ) d \xi \varepsilon \left( - q \cdot p \right)
 \delta \left( q^2 + 2q \cdot p \xi \right) \nonumber \\  \nonumber \\
 & = & \frac{1}{- q \cdot p} F ( \xi ) \nonumber
 \end{eqnarray}
 \\
 where function $F ( \xi )$ is $- q^2 / 2 q \cdot p$ and $F ( \xi )$
 is the Fourier transform of $\tilde{F} (p \cdot z)$:
 \begin{equation}
 F ( \xi ) = \frac{1}{2 \pi} \int e^{i \xi (p \cdot z)} \tilde{F}
 (p \cdot z) d (p \cdot z) \, .
 \end{equation}
 \\
 The function $F ( \xi )$ is therefore the Bjorken scaling function in
 the deep inelastic limit and is defined only for $- 1 < \xi < 1$ \, .
 We can write (3.6) in the form
 \begin{equation}
 F ( \xi ) = c_2 \cdot \delta ( \xi ) - c_4 \frac{1}{2!} \delta''( \xi )
 + c_6 \frac{1}{4!} \delta'''' (\xi ) - \cdots \, .
 \end{equation}
 The dimensionless numbers $c_i$ defined by the matrix elements of the
 expansion operators can be written as
 \begin{equation}
 c_2 = \int^1_{-1} F ( \xi ) d \xi, \qquad c_4 = - \int^1_{-1} F ( \xi )
 \xi^2 d \xi \cdots \, .
 \end{equation}
 This shows the connection between the matrix elements of the expansion
 operators and the moments of the scaling function. The Bjorken limit is
 seen to be a special case (the matrix element between single nucleon
 states of fixed momentum) of the light cone expansion.$^{17}$\\
 Now the derivation of the Bjorken limit from the light cone expansion
 can be described in terms of a kind of physical dimension $l$ for
 operators. (We shall see in the next section that these dimensions $l$
 are essentially the same as the physical dimensions $l$ we described
 in Section 2.) We define the expansion to conserve dimension on the light
 cone and assign to each current $l = - 3$ while counting each power of
 $z$ as having an $l$--value equal to the power. We see then that on the
 right--hand side we are assigning to each $J$--th rank Lorentz tensor
 (with
 maximum spin $J$) the dimension $l = - J - 2$. Furthermore, the physical
 dimension equals the mathematical dimension in all of these cases.\\
 \\
 \\
 {\bf 4. GENERALIZED LIGHT CONE SCALING AND BROKEN}\\
 {\bf SCALE INVARIANCE}\\
 \\
 We have outlined a situation in which scale invariance is broken by a
 non--vanishing $\Theta_{\mu \nu}$ but restored in the most singular terms
 of current commutators on the light cone. There is no reason to suppose
 that such a restoration is restricted to commutators of electromagnetic
 currents. We may extend the idea to all the vector currents
 ${\cal F}_{i \mu}$ and axial vector currents ${\cal F}_{i \mu} \, ^5$,
 to the scalar and pseudoscalar operators $u_i$ and $v_i$ that comprise
 the $\left( 3, \bar{3} \right)$ and $\left( \bar{3}, 3 \right)$
 representation thought to be involved in chiral symmetry breaking, to
 the whole stress--energy momentum tensor $\Theta_{\mu \nu} $, to any
 other local operators of physical significance, and finally to all the
local operators occurring in the light cone expansions of commutators
of all these quantities with one another. Let us suppose
 that in fact conservation of dimension applies to leading terms in the
 light cone in the commutators of all these quantities and that finally
 a closed algebraic system with an infinite number of local operators is
 attained, such that the light cone commutator of any two of the operators
 is expressible as a linear combination of operators in the algebra.
 We devote this section and the next one to discussing such a situation.\\
 If there is to be an analog of Bjorken scaling in all these situations,
 then on the right--hand side of the light cone commutation relations we
 want operators with $l = - J - 2$, as above for electromagnetic current
 commutators, so that we get leading matrix elements between one--particle
 states going like $c p_{\alpha} p_{\beta} \cdots$, where the $c$ are
 dimensionless constants.\\
 Of course, there might be cases in which, for some reason, all the
 $c$'s have to vanish, and the next--to--leading term on the light
 cone becomes the leading term. Then the coefficients would have the
 dimensions of positive powers of mass. We want to avoid, however,
 situations in which coefficients with the dimension of negative powers
 of mass occur; that means on the right--hand side we want
 $l \le - J - 2$ in any case, and $l = - J - 2$ when there is nothing
 to prevent it.\\
 This idea might have to be modified, as in a quark model with a scalar
 or pseudoscalar ``gluon'' field, to allow for a single operator $\phi $,
 with $l = - 1$ and $J = 0$, that can occur in a barred product, but
 without a sequence of higher tensors with $l = - J - 1$ that could
 occur in such a product; gradients of $\phi $ would, of course, average
 out in a barred product. However, even this modification is probably
 unnecessary, since preliminary indications are that, in the light cone
 commutator of any two physically interesting operators, the operator
 $\phi $ with $l = - 1$ would not appear on the right--hand side.\\
 Now, on the left--hand side, we want the non--conserved currents among
 ${\cal F}_{i \mu}$ and ${\cal F}_{i \mu}$$^5$ to act as if they have
 dimension $- 3$ just like the conserved ones, as far as leading
 singularities on the light cone are concerned, even though the
 non--conservation implies the admixture of terms that may have other
 dimensions $l$, dimensions that become $l - 1$ in the divergences, and
 correspond to dimensions $l - 1$ in the $SU_3 \times SU_3$ breaking
 terms in the energy density. But the idea of conservation of dimension
 on the light cone tells us that we are dealing with lower singularities
 when the dimensions of the operators on the left are greater. What is
 needed, then, is for the dimensions $l$ to be $> - 3$, i. e., for the
 chiral symmetry breaking terms in $\Theta_{\mu \nu}$ to have dimension
 $> - 4$. Likewise, if we want the stress--energy--momentum tensor itself
 to obey simple light cone scaling, we need to have the dimension of all
 scale breaking parts of $\Theta_{\mu \nu}$ restricted to values
 $> - 4$. In general, we can have symmetry on the light cone if the
 symmetry breaking terms in $\Theta_{\mu \nu}$ have dimension greater
 than $-4$. (See Appendix 1.)\\
 Now we can have ${\cal F}_{i \mu}$ and ${\cal F}_{i \mu}$$^5$ behaving,
 as far as leading singularities on the light cone are concerned, like
 conserved currents with $l = - 3, \Theta_{\mu \nu}$ behaving like a
 chiral and scale invariant quantity with $l = - 4$, and so forth. To
 pick out the subsidiary dimensions associated with the non--conservation
 of $SU_3 \times SU_3$ and dilation, we can study light cone commutators
 involving. $\partial_{\alpha} {\cal F}_{i \alpha}, \partial_{\alpha}
 {\cal F}_{i \alpha} \, ^5$, and
 $\Theta_{\mu \mu}$. (If the $\left( 3, \bar{3} \right) + \left( \bar{3},
 3 \right)$ hypothesis is correct, that means studying commutators
 involving $u$'s and $v$'s and also $\delta $, if $ \delta \not= 0.$\\
 In our enormous closed light cone algebra, we have all the operators
 under consi\-deration occuring on the left--hand side, the ones with
 $l = - J - 2$ on the right--hand side, and coefficients that are
 functions of $z$ behaving like powers according to the conservation of
 dimension. But are there restrictions on these powers? And are there
 restrictions on the dimensions occurring among the operators?\\
 If, for example, the functions of $z$ have to be like powers of $z^2$
 (or $\delta \left( z^2 \right), \delta' \left( z^2 \right)$ etc.)
 multiplied by tensors $z_{\alpha} z_{\beta} z_{\gamma} \cdots$, and if
 $l + J$ for some operators is allowed to be non--integral or even odd
 integral, then we cannot always have $l = - J - 2$ on the right, i. e.,
 the coefficients of all such operators would vanish in certain
 commutators, and for those commutators we would have to be content with
 operators with $l < - J - 2$
 on the right, and coefficients of leading tensors that act like positive
 powers of a mass.\\
 Let us consider the example:
 \begin{displaymath}
 \left[ \Theta_{\mu \nu} (x), u (y) \right] \hat{=} E_{\mu \nu} (z) \cdot
 \left( 0(y) + z_{\rho} O_{\rho} (y) + \cdots \right) + \cdots ,
 \end{displaymath}
 where $u(y)$ has the dimension $- 3$. In this case we cannot have the
 Bjorken scaling. Because of the relation
 \begin{displaymath}
 \left[ D (0), u(0) \right] = -3 i u (0),
 \end{displaymath}
 the operator $0(y)$ has to be proportional to $u(y)$. The operator series
 fulfilling the condition $l = - J - 2$ is forbidden in this case on the
 right--hand side.\\
 We have already emphasized that Nature seems to imitate the algebraic
 properties of free field theory rather than renormalized perturbation
 theory. (We could also say that Natrue is imitating a
 super--renormalizable theory, even though no sensible theory of that
 kind exists, with the usual methods of renormalization, in four
 dimensions.) This suggests that we should have in our general expansion
 framework finite equal--time commutators for all possible operators and
 their time derivatives.\\
 \\
 Such a requirement means that all functions of $z$ multiplying
 operators in light cone expansion must have the behavior described just
 above, i. e., the scalar functions involved behave like integral forces
 of $z^2$ or like derivatives of delta functions with $z^2$ as the argument.
 The formula
 \begin{displaymath}
 \frac{1}{\left( z^2 + i \varepsilon \right)^{\alpha}} -
 \frac{1}{\left( z^2 - i \varepsilon \right)^{\alpha}} \, _
 {\stackrel{\longrightarrow}{z_0 \rightarrow 0}}
 \, \, \, {\rm const.} \, \, \,
 z_0^{-2 \alpha + 3} \, \delta (z) 
 \end{displaymath}
 shows the sort of thing we mean. It also shows that $\alpha $ must not be
 too large. That can result in lower limits on the tensorial rank of the
 first operator in the light cone expansion in higher and higher tensors;
 to put it differently, the first few operators in a particular light
 cone expansion may have to be zero in order to give finiteness of
 equal time commutators with all time derivatives.\\
 Now, on the right--hand side of a light cone commutator of two
 physically interesting operators, when rules such as we have just
 discussed do not forbid it, we obtain operators with definite
 $SU_3 \times SU_3$ and other symmetry properties, of various tensor ranks,
 and with $l = - J - 2$. Now, for a given set of quantum numbers, how
 many such operators are there? Wilson$^1$ suggested a long time ago that
 there may be very few, sometimes only one, and others none. Thus no
 matter what we have on the left, we always would get the same old
 operators on the right (when not forbidden and less singular terms with
 dimensional coefficients occuring instead.) This is very important, since
 the matrix elements of these universal $l = - J - 2$ operators are then
 natural constants occurring in many problems. Wilson presumably went a
 little too far in guessing that the only Lorentz tensor operator in the
 light cone expansion of
 $\left[ \overline{j_{\mu} (x), j_{\nu}(y)} \right]$ would be the
 stress--energy--momentum tensor $\Theta_{\mu \nu}$, with no provision for
 an accompanying octet of $l = - 4$ tensors. That radical suggestion, as
 shown by Mack,$^{17}$ would make $\int F_2^{en} (\xi ) d \xi$ equal to
 $\int F_2^{ep} (\xi) d \xi$, which does not appear to be the case.
 However, it is still possible that one singlet and one octet of tensors
 may do the job. (See the discussion in Section 7 of the ``pure quark''
 case.)\\
 \\
 If we allow $z_0$ to approach zero in a light cone commutator, we obtain
 an equal time commutator. If Wilson's principle (suitably weakened) is
 admitted, then all physically interesting operator must obey some equal
 time commutation relations, with well--known operators on the right--hand
 side, and presumably there are fairly small algebraic systems to which
 these equal time commutators belong. The dimensions of the operators
 constrain severely the natur of the algebra involved. For example,
 suppose $SU_3 \times SU_3$ is broken by a quantity $u$ belonging to the
 representation $\left( 3, \bar{3} \right) \otimes
 \left( 3, \bar{3} \right)$ and having a singe dimension $l_u$. Then,
 if $l_u = - 3$, we may well have the algebraic system proposed years
 ago by one of us (M.G.--M.) in which $F_i$, $F_i^5$, $\int u_i d^3 x$
 and $\int v_i d^3 x$ obey the E.T.C. relations of $U_6$, as in the
 quark model. If $l_u = - 2$, however, then we would have $\int u_i d^3 x$
 and $d / dt \, \int u_i d^3 x$ commuting to give a set of quantities
 including $\int u_i d^3 x$ and so forth.\\
 We have described scaling in this section as if the dimensions $l$ were
 closely related to the dimensions $l$ obtained by equal time
 commutation with the dilation operator $D$ in Section 2. Let us now
 demonstrate that this is so.\\
 To take a simple case, suppose that in the light cone commutator of an
 operator $0 \cdots $ with itself, the same operator $0 \cdots $ occurs
 in the expansion on the right--hand side. Then we have a situation
 crudely described by the equation.
 \setcounter{page}{1}
\setcounter{equation}{0}
\renewcommand{\theequation}{4.\arabic{equation}}
 \begin{equation}
 \left[ O \cdots (z), O \cdots (0) \right] \hat{=} + (z)^l O \cdots
 (0) + \cdots \, ,
 \end{equation}
 where $l$ is the principal dimension of $O \cdots $. Here $(z)^l$ means
 any function of $z$ with dimension $l$, and we must have that because of
 conservation of dimension. Now under equal time commutation with $D$,
 say $O \cdots $ exhibits dimension $\bar l$. Let $z_0 \rightarrow 0$ and
 perform the equal time commutation, according to Eq. (2.3). We obtain
\begin{eqnarray}
\left( iz \cdot \bigtriangledown - 2 i \bar{l}  \right) \left[ O \ldots
(z), O \ldots (0) \right] & = & -i \bar{l} (z)^l O \ldots (0) \nonumber \\ \\
\left[ O \ldots (3), O \ldots (0) \right]
& = & \left( il - 2 i\bar{l} \right) (z)^l O \cdots (0) \nonumber
\end{eqnarray}
so that $l = \bar{l}$, as we would like.\\
Now to generalize the demonstration, we consider the infinite closed
algebra of light cone commutators, construct commutators like (4.1)
involving different operators, and from commutation with $D$ as in (4.2)
obtain equations
\begin{equation}
l_1 + l_2 - l_3 = \bar l_1  + \bar l_2  - \bar l_3 \, ,
\end{equation}
where $O \cdots^{(1)}$ and $O \cdots^{(2)}$ are commuted and yield a term
containing $O \cdots^{(3)}$ on the right. Chains of such relations can
then be used to demonstrate finally that $l = \bar{l}$ for the various
operators in which we are interested.\\
The subsidiary dimensions associated with symmetry breaking have not been
treated here. They can be dealt with in part by isolating the expressions
$\partial_{\mu} \, {\cal F}_{l \mu}^5, \Theta_{\mu \mu}$, etc., that
exhibit only the subsidiary dimensions and applying similar arguments to
them. In that way we learn that also for subsidiary dimensions
$\bar{l} = l$.\\
However, the subsidiary dimensions, while numerically equal for the two
definitions of dimension, do not enter in the same way for the two
definitions. The physical dimension $l$ defined by light cone commutation
always enters covariantly, while $l$ is defined by equal time commutation
with the quantity $D$ and enters non--covariantly, as in the break--up of
$\Theta_{\mu \nu}$ into the leading term $\stackrel{=}{\Theta}_{\mu \nu}$
of dimension
$ - 4$ and the subsidiary ones of higher dimensions. If these others
come from world scalars $w_n$ of dimensions $ \bar l_n$, then we have
we have\\
\begin{equation}
\Theta_{\mu \nu} = \stackrel{=}{\Theta}_{\mu \nu} + \sum\limits_{n}
\left\{ (3 + l) \delta _{\mu \nu} + (4 + l) \delta _{\mu 0} \,
\delta _{\nu 0} \right\} \frac{w_n}{3} \, ,
\end{equation}
so that we agree with the relations \\
\setcounter{page}{1}
\setcounter{equation}{7}
\renewcommand{\theequation}{2.\arabic{equation}}
\begin{eqnarray}
\Theta_{00} & = & \stackrel{=}{\Theta}_{00} + \sum\limits_{n} w_n \, ,
 \\ \nonumber \\
- \Theta_{\mu \mu} & = & \sum\limits_{n} \left( l_n + 4 \right) w_n \, .
\end{eqnarray}
Clearly, $\stackrel{=}{\Theta}_{\mu \nu}$ is non--covariant.\\
To obtain the non--covariant formula from the covariant one, the best
method is to write the light cone commutator of an operator with
$\Theta_{\mu \nu}$, involving physical dimensions $l$, and then construct
$D = - \int \, x_{\mu} \Theta_{\mu 0} d^3 x$ out of $\Theta_{\mu \nu}$
and allow the light cone commutator to approach an equal time
commutator. The non--covariant formula involving $\bar{l}$ must then
result.\\
As an example of non--covariant behavior of equal time commutation with
$D$, consider such a commutator involving an arbitrary tensor operator
$O_{\rho \sigma}$ of dimension $- 4$. We may pick up non--covariant
contributions that arise from lower order terms near the light cone
than those that give the dominant scaling behavior. We may have
\begin{displaymath}
\left[ \Theta_{\mu \nu} (x), 0_{\rho \sigma} (y) \right] =
\, \, \, {\rm leading \, \, \, term} \, \, \, + \partial_{\mu}
\partial_{\nu} \partial_{\rho} \partial_{\sigma} \left\{ \varepsilon
\left( z_0 \right) \delta \left( z^2 \right) \left[ O(y)
+ \cdots \right] \right\} + \cdots
\end{displaymath}
giving the result
\begin{displaymath}
{\rm E.T.C.} \, \, \, \left[ D, O_{\rho \sigma} (0) \right] = 4 i
\Theta_{\rho \sigma} (0) + \, \, \, {\rm const.} \, \, \, \delta_{\rho 0}
\delta_{\sigma 0} O(0) + \cdots \, .
\end{displaymath}
For commutation of $D$ with a scalar operator, there is no analog of this
situation.\\
\\
\\
{\bf 5. BILOCAL OPERATORS}\\
\\
So far, in commuting two currents at points separated by a
four--dimensional vector $z_{\mu}$, we have expanded the right--hand side on
the light cone in powers of $z_{\mu}$. It is very convenient for many
purposes to sum the series and obtain a single operator of low
Lorentz tensor rank that is a function of $z$. In a barred commutator,
it is a function of $z$ only, but in an ordinary unbarred commutator, it
is a function of $z$ and $R \equiv (x + y)/2$, in other words, a function
of $x$ and $y$. We call such an operator a bilocal operator and write
it as $O \cdots (x, y)$ or, in barred form, $\bar{O} \cdots (x, y)$.\\
We can, for example, write Eq. (3.3) in the form
\setcounter{page}{1}
\setcounter{equation}{0}
\renewcommand{\theequation}{5.\arabic{equation}}
\begin{equation}
\overline{\left[ j_{\mu} (x), j_{\nu}(y) \right]} \hat{=}
t_{\mu \nu \rho \sigma} \left\{ \varepsilon \left( z_0 \right)
\delta \left( z^2 \right) \overline{O_{\rho \sigma}(x, y)} \right\} +
\, \, \, {\rm longitudinal \, \, \, term},
\end{equation}
using the barred form of a bilocal operator $O_{\rho \sigma} (x, y)$
that sums up all the tensors of higher and higher rank in Eq. (3.3).\\
Now in terms of bilocal operators we can formulate a much stronger
hypothesis than the modified Wilson hypothesis mentioned in the last
section. There we supposed that on the right--hand side of any
light--cone commutators (unless the leading terms were forbidden for some
reason) we would always have operators with $l = - J - 2$ and that for a
given $J$ and a given set of quantum numbers there would be very few of
these, perhaps only one, and that the quantum numbers themselves would
be greatly restricted (for example, to $SU_3$ octets and singlets). Here
we can state the much stronger conjecture that for a given set of quantum
numbers the bilocal operators appearing on the right are very few in
number (and perhaps there is only one in each case), with the quantum
numbers greatly restricted. That means that instead of an arbitrary
series
$O_{\rho \sigma} + \, \, \ {\rm const.} \, O_{\rho \sigma \lambda \mu}
\, 'z_{\lambda} z_{\mu} z_{\alpha} z_{\beta}
O_{\alpha \beta \rho \sigma \lambda \mu} + \cdots $, we have a unique sum
$O_{\rho \sigma} (x, y)$ with all the
constants determined. The same bilocal operator will appear in many
commutators, then, and its matrix elements (for example, between proton
and proton with no change of momentum) will give universal deep inelastic
form factors.\\
Let us express in terms of bilocal operators the idea mentioned in the
last section that all tensor operators appearing on the right--hand side
of the light cone current commutators may themselves be commuted according
to conservation of dimension on the light cone, but lead to the same set
of operators, giving a closed light cone algebra of an infinite number of
local operators of all tensor ranks. We can sum up all these operators
to make bilocal operators and commute those, obtaining, on the
right--hand side according to the principle mentioned above, the same
bilocal operators. Thus we obtain a light cone algebra generated by a
small finite number of bilocal operators. These are the bilocal operators
that give the most singular terms on the light cone in any commutator
of local operators, the terms that give scaling behavior. (As we have
said, in certain cases they may be forbidden to occur and positive powers
of masses would then appear instead of dimensionless coefficients.)\\
This idea of a universal light cone algebra of bilocal operators with
$l = - J - 2$ is a very elegant hypothesis, but one that goes far beyond
present experimental evidence. We can hope to check it some day if we
can find situations in which limiting cases of experiments involve the
light cone commutators of light cone commutators. Attempts have been
made to connect differential cross sections for the Compton effect with
such mathematical quantities;$^5$ it will be interesting to see what
comes of that and other such efforts.\\
A very important technical question arises in connection with the light
cone algebra of bilocal operators. When we talk about the commutators of
the individual local operators of all tensor ranks, we are dealing with
just two points $x$ and $y$ and with the limit $(x - y)^2 \rightarrow
0$. but when we treat the commutator of bilocal operators
$O (x, u)$ and $O(y, v)$, what are the space--time relationships of
$x, u, y$, and $v$ in the case to which the commutation relations apply?
We must be careful, because if we give too liberal a prescription for
these relationships we may be assuming more than could be true in any
realistic picture of hadrons.\\
The bilocal operators arise originally in commutators of local operators
on the light cone, and therefore we are interested in them when
$(x - u)^2 \rightarrow 0$ and $(y - v)^2 \rightarrow 0$. In the light
cone algebra of bilocal operators, we are interested in singularities
that are picked up when $(x - y)^2$ or when $(u - v)^2 \rightarrow 0$ or
when $(x - v)^2 \rightarrow 0$ or when $(u - y)^2 \rightarrow 0$. But
do we have to have all six quantities simultaneously brought near to
zero? That is not yet clear. In order to be save, let us assume here that
all six quantities do got to zero.\\
\\
\\
{\bf 6. LIGHT CONE ALGEBRA ABSTRACTED FROM A}\\
{\bf QUARK PICTURE}\\
\\
Can we postulate a particular form for the light cone algebra of bilocal
operators?\\
\\
We have indicated above that if the Stanford experiments, when
extended and refined, still suggest the absence of logarithmic terms, the
vanishing of the longitudinal cross section, and a difference between
neutron and proton in the deep inelastic limit, then it looks as if in
this limit Nature is following free field theory, or interacting field
theory with naive manipulation of operators, rather than what we know
about the perturbation expansions of renormalised field theory. We might,
therefore, look at a simple relativistic field theory model and abstract
from it a light cone algebra that we could postulate as being true of
the real system of hadrons. The simplest such model would be that of free
quarks.\\
In the same way, the idea of an algebra of equal--time commutators of
charges or charge densities was abstracted ten years ago from a
relativistic Lagrangian model of a free spin 1/2 triplet, what would
nowadays be called the quark triplet. The essential feature in this
abstraction was the remark that turning on certain kinds of strong
interaction in such a model would not affect the equal time commutation
relations, even when all orders of perturbation theory were included;
likewise, mass differences breaking the symmetry under $SU_3$ would not
disturb the equal time commutation relations of $SU_3$.\\
We are faced, then, with the following question. Are there non--trivial
field theory models of quarks with interactions such that the light cone
algebra of free quarks remains undisturbed to all orders of naive
perturbation theory? Of course, the interactions will make great changes
in the operator commutators inside the light cone; the question is
whether the leading singularity on the light cone is unaffected. Let
us assume, for purposes of our discussion, that the answer is affirmative.
Then we can feel somewhat safe from absurdity in postulating for real
hadrons the light cone algebras of free quarks, and indeed of massless
free quarks (since the masses do not affect the light cone singularity).\\
Actually, it is easy to construct an example of an interacting field
theory in which our condition seems to be fulfilled, namely a theory in
which the quark field interacts with a neutral scalar or pseudoscalar
``gluon'' field $\phi$. We note the fact that the only operator series
in such a theory that fulfills $l = - J - 2$ and contains $\phi (x)$ is
the following: $\phi(x)\phi(x), \phi(x)\partial_{\mu}\phi(x) \cdots$.
But these operators do not seem to appear in light cone expansions of
products of local operators consisting only of quark fields, like the
currents. A different situation prevails in a theory in which the
``gluon'' is a vector meson, since in that case we can have the
operator series
$\bar{q}(x)\gamma_{\mu}B_{\nu}(x)\gamma(x), \bar{q}(x)\gamma_{\mu}
B_{\nu} B_{\rho}q(x), \cdots $, contributing to the Bjorken limit. The
detailed behavior of the various ``gluon'' models is being studied by
Llewellyn Smith.$^{18}$.\\
\newpage
\noindent
In the following, we consider the light cone algebra suggested by the
quark model. We obtain for the commutator of two currents on the light
cone (connected part only):
\setcounter{page}{1}
\setcounter{equation}{0}
\renewcommand{\theequation}{6.\arabic{equation}}
\begin{eqnarray}
\left[ {\cal F}_{i \mu}(x), {\cal F}_{j \nu} (y) \right]
& \hat{=} & \frac{1}{4 \pi} \partial_{\rho} \left[ \varepsilon
\left( z_0 \right) \delta \left( z^2 \right) \right] \left\{ i f_{ijk}
\left[ S_{\mu \nu \rho \sigma} \left( {\cal F}_{k \sigma} (x, y) +
{\cal F}_{k \sigma} (y, x) \right) \right. \right.
\nonumber \\ \nonumber\\
& & \left. + i \varepsilon_{\mu \nu \rho \sigma}
\left( {\cal F}^5 _{k \sigma} (y, x) - {\cal F}^5 _{k \sigma}
(x, y) \right) \right] + d_{ijk}
\left[ s_{\mu \nu \rho \sigma} \left( {\cal F}_{k \sigma} (x, y)
\right. \right. \nonumber \\  \nonumber \\
& & \left. \left. \left. \left. - \left( {\cal F}_{k \sigma} (y, x) \right)
- i \varepsilon_{\mu \nu \rho \sigma} \left( {\cal F}^5_{k \sigma}
(y, x) \right) + {\cal F}^5_{k \sigma} (x, y) 
\right) \right) \right] \right\} \, , \nonumber
\end{eqnarray}
\\
\begin{eqnarray}
\left[ {\cal F}^5_{i \mu}(x), {\cal F}_{j \nu} (y) \right]
& \hat{=} & \frac{1}{4 \pi} \partial_{\rho}
\left[ \varepsilon \left( z_0 \right) \delta \left( z^2 \right) \right]
\left\{  i f _{ijk} \left[ s_{\mu \nu \rho \sigma}
\left( {\cal F}^5_{k \sigma} (x, y) + {\cal F}^5_{k \sigma}
\left( y (x) \right) \right. \right. \right. \nonumber \\  \nonumber \\
& &  + i \varepsilon_{\mu \nu \rho \sigma}
\left( {\cal F}_{k \sigma} (y, x) - {\cal F}_{k \sigma}
(x, y)   \right)  \nonumber \\ \nonumber \\
& & + d_{ijk} \left[ s_{\mu \nu \rho \sigma}
\left( {\cal F}^5_{k \sigma} (x, y) \right)
 - \left( {\cal F}^5_{k \sigma} (y, x) \right) \right] \nonumber \\
\\
& & \left. \left. \left. - i \varepsilon_{\mu \nu \rho \sigma}
\left( {\cal F}^5_{k \sigma} (x, y) + {\cal F}_{k \sigma} (x, y) \right)
\right) \right] \right\} \nonumber \\  \nonumber \\
& & \left[ {\cal F}^5_{i \mu} (x), {\cal F}^5 \, _{j \nu} (y) \right]
= \left[ {\cal F}_{i \mu} (x), {\cal F}_{j \nu} (y) \right] \, ,
\nonumber \\  \nonumber \\
s_{\mu \nu \rho \sigma} & = & \delta_{\mu \sigma} \delta_{\nu \sigma} +
\delta_{\nu \rho} \delta_{\mu \sigma} - \delta_{\mu \nu}
\delta_{\rho \sigma} \, , \qquad z = x - y. \nonumber
\end{eqnarray}
If we go to the equal time limit in (6.1) we pick up the current algebra
relations for the currents; in fact we obtain, for the space integrals of
all components of nine vector and nine axial--vector currents, the
algebra$^{19}$ of $U_6 \times U_6$.\\
Note that we can get similar relations for the current anti--commutators
or for the products of currents on the light cone, just be replacing
\begin{displaymath}
\frac{1}{4 \pi} \partial_{\rho} \left[ \varepsilon \left( z_0 \right)
\delta \left( z^2 \right) \right] \, \, \, {\rm by} \, \, \,
- \frac{i}{4 \pi^2} \partial_{\rho} \frac{1}{z^2} \, \, \, {\rm or
\, \, \, by} \, \, \, - \frac{i}{8 \pi^2} \partial_{\rho}
\frac{1}{z^2 + i \varepsilon z_0}
\end{displaymath}
respectively. Perhaps we can abstract these relations also and use them
for hadron theory.\\
\newpage
\noindent
In (6.1) we have introduced bilocal generalizations of the vector and
axial--vector currents, which in a quark model correspond to products
of quark fields:
\begin{eqnarray}
{\cal F}_{k \sigma} (x, y) & \sim & \tilde{q}(x) \frac{i}{2} \lambda_k
\gamma_{\sigma} q (y), \nonumber \\    \\
{\cal F}^5_{k \sigma} (x, y) & \sim & \tilde{q}(x) \frac{i}{2} \lambda_k
\gamma_{\sigma} \gamma_5 q (y) \, . \nonumber
\end{eqnarray}
Note that the products in (6.2) have to be understood as ``generalized
Wick products''. The $c$--number part in the product of two quark fields
is already excluded, since it does not contribute to the connected current
commutator. The $c$--number part is measured by vacuum processes like
$e^+ e^-$ annihilation. Assuming that the disconnected part of the
commutator on the light cone is also dictated by the quark model, we
would obtain
\begin{displaymath}
\sigma_{{\rm tot} \, \, \, e^+ e^-} \sim \, \, \, {\rm const./s \, \, \,
for} \, \, \, e^+ e^- \, \, \, {\rm annihilation} \, ,
\end{displaymath}
where $s$ is as usually defined: $s = - \left( p_1 + p_2 \right)^2$. In
particular, we would get
\begin{displaymath}
\sigma_{{\rm tot}} \left( e^+ e^- \, \, \, {\rm into \, \, \, hadrons}
\right) \rightarrow \left( \sum Q^2 \right) \sigma_{{\rm tot}} \, \,
\left( e^+ e^- \, \ \ {\rm into \, \, \, muons} \right)
\end{displaymath}
with $\sum Q^2  = (2/3)^2 + (1/3)^2 = 2/3$.\\
Now we go on to close the algebraic system of (6.1), where local currents
occur on the left--hand side and bilocal ones on the right.\\
Let us assume that the bilocal generalizations of the vector and axail
vector currents are the basic entities of the scheme. Again using the
quark model as a guideline on the light cone, we obtain the following
closed algebraic system for these bilocal operators:\\
\begin{eqnarray}
& & \left[ {\cal F}_{i \mu} (x, u), {\cal F}_{j \nu}(y, v) \right] 
\nonumber \\ \nonumber \\
& \hat{=} & \frac{1}{4 \pi} \partial_{\rho} \left\{ \varepsilon
\left( x_0 - v_0 \right) \delta \left[ (x - v)^2  \right] \right\}
\left( i f_{ijk} - d_{ijk} \right) \left( s_{\mu \nu \rho \sigma}
{\cal F}_{k \sigma} (y, u) \right. \nonumber \\  \nonumber \\
& & \left. + i \varepsilon_{\mu \nu \rho \sigma} {\cal F}^5_{k \sigma}
(y, u) \right) + \frac{1}{4 \pi} \partial_{\rho} \left\{ \varepsilon
\left( u_0 - y_0 \right) \partial \left[ (u - y)^2 \right] \right\}
\left( if_{ijk} + d_{ijk} \right) \nonumber \\  \nonumber \\
& & \cdot \left( s_{\mu \nu \rho \sigma} {\cal F}_{k \sigma}
(x, v) - i \varepsilon_{\mu \nu \rho \sigma} {\cal F}^5_{k \sigma}
(x, v) \right), \nonumber \\
\end{eqnarray}
\newpage
\noindent
\begin{eqnarray}
& &  \left[ {\cal F}^5_{i \mu} (x, u), {\cal F}_{j \nu} (y, v) \right] 
 \nonumber \\ \nonumber \\
& \hat{=} & \frac{1}{4 \pi} \partial_{\rho} \left\{ \varepsilon
\left( x_0 - v_0 \right) \partial \left[ (x - v)^2 \right] \right\}
\left( if_{ijk} - d_{ijk} \right) \nonumber \\  \nonumber \\
& & \left( s_{\mu \nu \rho \sigma} {\cal F}^5_{k \sigma}(y, u) + i
\varepsilon_{\mu \nu \rho \sigma} {\cal F}_{k \sigma} (y, u) \right)
\nonumber \\  \nonumber \\
& & + \frac{1}{4 \pi} \partial_{\rho} \left\{ \varepsilon
\left( u_0 - y_0 \right)
\delta \left[ (u - y)^2 \right]  \right\} \left( if_{jik} + d_{ijk}
\right) \nonumber \\  \nonumber \\
& &  \cdot \left( s_{\mu \nu \rho \sigma} {\cal F}^5_{k \sigma} (x, y) - i
\varepsilon_{\mu \nu \rho \sigma} {\cal F}{k \sigma} (x, v)  \right),
\nonumber \\  \nonumber\\
& & \left[ {\cal F}^5_{i \mu} (x, u), {\cal F}^5_{j \nu} (y, v) \right]
\hat{=} \left[ {\cal F}_{i \mu} (x, u), {\cal F}_{j \nu} (y, v)
\right] \, . \nonumber
\end{eqnarray}
Similar relations might be abstracted for the anticommutators and products
of two bilocal currents near the light cone. The relations (6.3) are
assumed to be true if
\begin{eqnarray}
(x - u)^2 \approx 0, & \, & (u - y)^2 \approx 0, \nonumber \\
(u - v)^2 \approx 0, & \, & (x - y)^2 \approx 0, \nonumber \\
(x - v)^2 \approx 0, & \, & (u - v)^2 \approx 0   \nonumber \, .
\end{eqnarray}
This condition is obviously fulfilled if the four points $x, u, y, v$ are
distributed on a straight line on the light cone. The algebraic relations
(6.3) can be used, for example, to determine the light cone commutator of
two light cone commutators and relate this more complicated case to the
simpler case of a light cone commutator. It would be interesting to
propose experiments in order to test the relations (6.3).\\
\\
\\
{\bf 7. LIGHT CONE ALGEBRA AND DEEP INELASTIC SCATTERING}\\
\\
In the last section we have emphasized that perhaps the light cone is a
region of very high symmetry (scale and $SU_3 \times SU_3$ invariance).
Furthermore, we have abstracted from the quark model certain albebraic
properties that might be right on the light cone. Now we should like to
mention some general relations that we can obtain using this light cone
algebra. But let us first consider the weak interactions in the deep
inelastic region.
\newpage
\noindent
We introduce the weak currents $J _{\mu}^+ (x), J_{\nu}^-(x)$ and
consider the following expression:
\setcounter{page}{1}
\setcounter{equation}{0}
\renewcommand{\theequation}{7.\arabic{equation}}
\begin{eqnarray}
W_{\mu \nu}(q) & = & \frac{1}{4 \pi} \int d^4 ze^{-iq\cdot z} < \mid
\left[ J_{\mu}^+ (z)  , J_{\nu}^- (0) \right] \mid p > \nonumber \\
\nonumber \\
& = & \left( \delta_{\mu \nu} - \frac{q_{\mu} q_{\nu}}{q^2} \right)
\left( W_1 \, ^+ - \frac{p \cdot q}{q^2} W_2^+ \right) - \frac{i}{2}
\varepsilon_{\mu \nu \alpha \beta} p_{\alpha} p_{\beta} W_3^+
\nonumber \\  \nonumber \\
& & \frac{\delta_{\mu \nu} (p \cdot q)^2 + p_{\mu} p_{\nu} q^2 -
\left( p_{\mu } p_{\nu} + p_{\nu} p_{\mu} \right) p \cdot q}{q^2}
W_2^+ + q^{\mu} q^{\nu} W_4^+ \nonumber \\  \nonumber \\
& & + \left( q_{\mu} p_{\nu} + q_{\nu} p_{\mu} \right) W_5^+ + i
\left( q_{\mu} p_{\nu} + q_{\nu} p_{\mu} \right) W_6 \, .
\end{eqnarray}

In general, we have to describe the inelastic neutrino hadron processes by
six structure functions. From naive scaling arguments we would expect
in the deep inelastic limit:
\begin{eqnarray}
W_1 \, ^+ \rightarrow F_1 (\xi ), & & - q \cdot p W_2 \, ^+ \rightarrow
F_2 (\xi), \nonumber \\ \nonumber \\
- q \cdot p W_3^+ \rightarrow F_3 (\xi), & & - q \cdot p W_4 \, ^+
\rightarrow F_4 (\xi), \\
\nonumber \\
- q \cdot p W_5 \, ^+ \rightarrow F_5 (\xi ), & & - q \cdot p W_6 \, ^+
\rightarrow F_6 (\xi). \nonumber
\end{eqnarray}

The formulae above have the most general form, valid for arbitrary vectors
$J_{\mu}(x)$. We neglect the $T$--violating effects, which may in any
case be $0$ on the light cone: $F_6 = 0$. We have already stressed that the
weak currents are conserved on the light cone, and we conclude:
\begin{equation}
F_4 (\xi) = F_5 (\xi) = 0.
\end{equation}

Equation (7.3) is an experimental consequence of the $SU_3 \times SU_3$
symmetry on the light cone, which may be tested by exerpiment. In the
deep inelastic limit we have only three non--vanishing structure
functions, corresponding to a conserved current.\\
It is interesting to note that there is the possibility of testing the
dimension $l$ of the divergence of the axial vector current, if our
scaling hypothesis is right. We write, for the weak axial vector current,
\begin{equation}
\partial_{\mu} {\cal F}^5_{\pm \mu} = c \cdot v_{\pm}(x)
\end{equation}
where $v_{\pm} (x)$ is a local operator of dimension $l$. and $c$ is a
parameter with non--zero dimension.\\
\newpage
\noindent
According to our assumptions about symmetry breaking, $c$ can be written
as a positive power of a mass. Using (7.1), we obtain
\begin{eqnarray}
q^{\mu} q^{\nu} W_{\mu \nu}\, ^+ (q) & = & \frac{c^2}{4 \pi} \int d^4
z e^{- iq \cdot z} < p \mid \left[ v_+ (z), v_- (0) \right] \mid p >
 \\  \nonumber 
\\
& = & \left( q^2 \right)^2 W_4 \, ^+ - 2 q^2 q \cdot p W_5 \, ^+ \, .
\end{eqnarray}
\\
We define:
\begin{equation}
D \left( q^2, q \cdot p \right) = \frac{1}{4 \pi} \int d^4 ze^{-iq \cdot z}
< p \mid \left[ v_+(z), v_-(0) \right] \mid p >.
\end{equation}

If we assume that $D$ scales in the deep inelastic region according to
the dimension $l$ of $v_{\pm}(x)$, we obtain
\begin{equation}
\lim_{bj} ( - p \cdot q)^{-l-3} D \left( q^2, q \cdot p \right) =
\phi (\xi)
\end{equation}
where $\phi (\xi)$ denots the deep inelastic structure function for the
matrix element (7.6).\\
Using (7.5) we obtain
\begin{equation}
\lim_{bj} (- p \cdot q)^{5+l} \left( \xi^2 W_4 \, ^+ - 2 \xi W_5 \,
^+ \right) = c^2 \phi (\xi) \, .
\end{equation}
If we determine experimentally the scaling properties of $W_4$ and $W_5$,
then we can deduce from (7.8) the dimension $l$ of $v_{\pm} (x)$. This
$l$ is the same quantity as the dimension $\bar{l}_u$ discussed in
Section 2, provided the $SU_3 \times SU_3$ violating term in the energy
has a definite dimension.$^{20}$\\
In order to apply the light cone algebra of Section 6, we have to relate
the expectation values of the bilocal operators appearing there to
the structure function in question. This is done in Appendix II, where
we give this connection for arbitrary currents. We use Eqs. (A.12) and
(A.13), where the functions $S^k (\xi), A^k (\xi)$ are given by the
expectation value of the symmetric an antisymmetric bilocal currents
(Eq.(A.8)), and obtain:
\begin{enumerate}
\item[(a)] for deep inelastic electron--hadron scattering:
\begin{equation}
F^{ep}_2 (\xi ) = \xi \left( \frac{2}{3} \sqrt{\frac{2}{3}} A^0 ( \xi)
+ \frac{1}{3 \sqrt{3}} \, A^8 (\xi) + \frac{1}{3} A^3 (\xi) \right)
\end{equation}

\item[(b)] for deep inelastic neutrino--hadron scattering:
\begin{eqnarray}
F_2^{\nu p} (\xi) & = & \xi \left( 2S^3 (\xi) + 2 \sqrt{\frac{2}{3}} A^0
(\zeta) + \frac{2}{\sqrt{3}} A^8 (\xi) \right)  \\  \nonumber \\
F_3^{\nu p} (\xi) & = & 2 A^3 (\xi) - 2 \sqrt{\frac{2}{3}} S^0
(\xi) - \frac{2}{\sqrt{3}} S^8 (\xi) \, .
\end{eqnarray}
\end{enumerate}
In (7.5) and (7.6) we have neglected the Cabibbo angle, since
${\rm sin}^2 \Theta_c = 0.05 \approx 0$.\\
Both in (7.4) and (7.6), $A^3 (\xi)$ occurs as the only isospin dependent
part, and we can simply derive relations between the structure functions
of different members of an isospin multiplet, e. g., for neutron and
proton:
\begin{equation}
6 \cdot \left( F^{en}_2 - F_2^{ep}  \right) = \xi \cdot
\left( F_3^{\nu p} - F^{\nu n}_3 \right) \, .
\end{equation}
This relation was first obtained by C. H. Llewellyn Smith$^7$ within
the ``parton'' model. One can derive similar relations for other
isospin multiplets.\\
In the symmetric bilocal current appear certain operators that we know.
The operator $j_{\mu}(x) = i \bar{q}(x)y_{\mu}q(x)$ has to be identical
with the hadron current (we suppress internal indices) in order to give
current algebra. But we know their expectation values, which are given by
the corresponding quantum number. In such a way we can derive a large
set of sum rules relating certain moments of the structure functions to
their well--known expectation values.\\
We give only the following two examples, which follow immediately from
(7.10), (7.11), (7.12):
\begin{eqnarray}
\int^1_{-1} \frac{d \xi}{\xi} \left( F^{\nu p}_2 (\xi) - F^{\nu n}_2
(\xi) \right) & = & \int^1_{-1} \frac{d \xi}{\xi} \left( F^{\nu p}_2
(\xi ) - F^{\nu p}_2 \left( - \xi \right) \right) \nonumber \\
 \\
& = & 4 s_1^3 (p) = 4. \nonumber
\end{eqnarray}
Here $s_1^3 (p)$ means, as in Appendix II, the proton expectation value
of $2 F_3$. This is the Adler sum rule,$^{21}$, usually written as
\begin{equation}
\int^1_0 \frac{d \xi}{\xi} \left( F^{\nu p}_2 (\xi) - F^{\nu n}_2
(\xi) \right) = 2.
\end{equation}

From (7.11) we obtain:
\begin{equation}
\int\limits_{-1}^1 \, \left( F_3^{\nu p} + F_3^{\nu n} \right) d \xi = - 2
\left( 2s_1^0 (p) + s_1^8 (p) \right) = - 12
\end{equation}
or
\begin{equation}
\int\limits_{0}^1 \left( F_3^{\nu p} + F_3^{\nu n} \right) d \xi = - 6,
\end{equation}
which is the sum rule first derived by Gross and Llewellyn Smith.$^{22}$\\
If we make the special assumption that we are abstracting our light
cone relations from a pure quark model with no ``gluon field'' and
non--derivative couplings, we can get a further set of relations.\\
Of course, no such model is known to exist in four dimensions that is
even renormalizable, much less super--renormalisable as we would prefer
to fit in with the ideas presented here. Nevertheless, it may be
worthwhile to examine sum rules that test whether Nature imitates the
``pure quark'' case.\\
The point is that when we expand the bilocal quantity
${\cal F}_{0 \alpha}(x,y)$
to first order in $y - x$, we pick up a Lorentz tensor operator, a singlet
under $SU_3$, that corresponds in the quark picture to the operator
$1/2 \left\{ \bar{q}(x)\gamma_{\mu} \partial_{\nu} q (x)
- \partial_{\nu} \bar{q}(x)\gamma_{\mu} q(x) \right\}$, which, if we
symmetrize in $\mu $ and $\nu $ and ignore the
trace, is the same as the stress--energy--momentum tensor
$\Theta_{\mu \nu}$ in the pure quark picture. But the expected value of
$\Theta_{\mu \nu}$ in any state of momentum $p$ is just
$2 p_{\mu} p_{\nu}$, and so we obtain sum rules for the pure quark case.\\
We consider the isospin averaged expressions:
\begin{eqnarray}
\left( F_2^{ep} (\xi ) + F_1^{en} (\xi) \right) & = & 2\xi \left\{
\frac{2}{3} \sqrt{\frac{2}{3}} A^0 (\xi) + \frac{1}{3} \frac{1}{\sqrt{3}}
A^8 (\xi)  \right\} \nonumber \\ \nonumber \\
\left( F^{\nu p}_2 (\xi) + F^{\nu n}_2 (\xi) \right) & = & 2 \xi \left\{
2 \sqrt{\frac{2}{3}} A^0 (\xi) + \frac{2}{\sqrt{3}} A^8 (\xi)  \right\}
\nonumber
\end{eqnarray}
\\ and obtain \\
\begin{displaymath}
6 \left( F^{ep}_2 + F^{en}_2 \right) - \left( F^{\nu p}_2 + F^{\nu n}_2
 \right) = 4 \sqrt{ \frac{2}{3}} A^0 (\xi)
\end{displaymath}
\begin{displaymath}
= 4 \sqrt{\frac{2}{3}} \left( a^0_1 \delta (\xi) - \frac{1}{2!} a^0_3
\delta'' (\xi) \cdots \right)
\end{displaymath}
\\
In pure quark theories we have $a^0_1 = \sqrt{2/3}$ and we obtain
\begin{displaymath}
6 \int^1_{-1} \left( F^{ep}_2 + F^{en}_2 \right) d (\xi) - \int^1_{-1}
\left( F^{\nu p}_2 (\xi) + F^{\nu n}_2 (\xi) \right) d \xi = 8/3
\end{displaymath}
\\
or, for the physical region $0 \le \xi \le 1$:
\begin{equation}
6 \int^1_0 \left( F^{ep}_2 + F^{en}_2 \right) d \xi - \int^1_0
\left( F^{\nu p}_2 + F^{\nu n}_2 \right) e \xi = 4/3 \, .
\end{equation}
\\
The sum rule (7.18) can be tested by experiment. This will test whether
one can describe the real world of hadrons by a theory resembling one
with only quarks, interacting in some unknown non--linear fashion.\\
The scaling behavior in the deep inelastic region may be described by the
``parton model''. $^{4, 5}$ In the deep inelastic region, the electron is
viewed as scattering in the impulse approximation off point--like
constituents of the hadrons (``partons''). In this case the scaling function
$F^e_2 (\xi)$ can be written as
\begin{equation}
F^e_2 (\xi) = \sum\limits_{N} P(N) \, \left( \sum\limits_{i} \, Q_i \, ^2
\right)_N \xi \, f_N (\xi)
\end{equation}
where we sum up over all ``partons'' $\left( \sum _i \right)$ and all the
possibilities of having $N$ partons $\left( \sum_N \right)$. The momentum
distribution function of the ``partons'' is denoted by $f_8 (\xi)$, the
charge of the i-th ``partin'' by $Q_i$. We compare (7.9) with (7.18):
\begin{eqnarray}
F^e_2 (\xi) & = & \xi \left( \frac{2}{3} A^0 (\xi) + \frac{1}{6} A^8 (\xi)
+ \frac{1}{3} A^3 (\xi) \right) \nonumber \\  \\
& = & \sum\limits_N P(N) \left( \sum\limits_i \left( Q_i \, ^2 \right)_N
\xi f_N (\xi). \right) \nonumber
\end{eqnarray}
\\
As long as we do not specify the functions $f_N(\xi)$ and $P(N)$, the
``parton model'' gives us no more information than the generalization of
current algebra to the light cone as described in the last sections. If one
assumes special properties of these functions, one goes beyond the light
cone algebra of the currents, that means beyond the properties of the
operator products on the light cone. Such additional assumptions. e. g.,
statistical assumptions about the distributions of the ``partons'' in
relativistic phase space, appear in the light cone algebra approach as
specific assumptions about the matrix elements of the expansion operators
on the light cone. These  additional assumptions are seen, in our approach,
to be model dependent and somewhat arbitrary, as compared to results of
the light cone algebra. Our results can, of course, be obtained by
``parton'' methods and are mostly well--known in that connection.\\
\\
It is interesting to consider the different sum rules within the
``parton model''. The sum rules (7.15) and (7.17) are valid in any
``quark--parton'' model; so is the symmetry relation (7.18). The sum rule
(7.18) is a specific property of a model consisting only of quarks. If
there is a ``gluon'' present, we obtain a deviation from 4/3 on the
right--hand side, which measures the ``gluon'' contribution to the
energy--momentum tensor.\\
\\
Our closed algebra of bilocal operators on the light cone has, of course,
a parallel in the ``parton'' model. However it is again much easier using
our approach to disentangle what may be exactly true (formulae for light
cone commutators of light cone commutators) from what depends on specific
matrix elements and is therefore model dependent. It would be profitable
to apply such an analysis to the work of Bjorken and Paschos, in the
context of ``partons'', on scaling in the Compton effect on protons.\\
\\
As an example of a ``parton model'' relation that mingles specific
assumptions about matrix elements with more general ideas of light cone
algebra and abstraction from a pure quark model, we may take the allegation
that in the pure quark case we have $\int F^{en}_2 (\xi) d \xi = 2/9$.
Light cone algebra and the pure quark assumption do not imply this.\\
\\
\\
{\bf 8. CONCLUDING REMARKS}\\
\\
There are many observations that we would like to make and many
unanswered questions that we would like to raise about light cone algebra.
But we shall content ourselves with just a few remarks.\\
\\
First comes the question of whether we can distinguish in a well--defined
mathematical way, using physical quantities, between a theory that makes
use of $SU_3$ triplet representations locally and one that does not. If
we can, we must then ask whether a theory that has triplets locally
necessarily implies the existence of real triplets (say real quarks)
asymptotically. Dashen (private communication) raises these two questions
by constructing local charge operators $\int_V {\cal F}_{i0} d^3x$ over
a finite volume. (This construction is somewhat illegitimate, since test
functions in field theory have to be multiplied by $\delta $ functions
in equal time charge density commutators and should therefore have all
derivatives, not like the function that Dashen uses, which is unity inside
$V$ and zero outside.) If his quantities $F_i^V$ make sense, they obey
the commutation rules of $SU_3$ and we can ask whether for any $V$ our
states contain triplet (or other triality $\not= 0$) representation of
this $SU_3$. Dashen then suggests that our bilocal algebra probably implies
that local triplets in this sense are present; if the procedure and the
conclusion are correct, we must ask whether real quarks are then implied.\\
\\
The question of quark statistics is another interesting one. If quarks
are real, then we cannot assign them para--Fermi statistics of rank 3,
since that is said to violate the factoring of the $S$--matrix for
distant subsystems. However, if somehow our quarks are permanently bound
in oscillators (and our theory is thus perhaps equi\-valent to a bootstrap
theory with no real quarks), then they could be parafermions of rank 3.
They can be bosons, too, if they are not real, but only if there is a
spinless fermion (the ``soul'' of a baryon) that accompanies the three
quarks in each baryon.\\
\\
Another topic is the algebra of $U_6 \times U_6 \times O_3$ that is
implied at equal times for the integrals of the current component and
the angular momentum.$^{19}$ Is that algebra really correct or is it
too strong an assumption? Should it be replaced at $P_s = \infty$ by
only the ``good--good'' part of the algebra?\\
\\
If we do have the full algebra, then the quark kinetic part of the
energy density is uniquely defined as the part behaving like ({\bf 35, 1})
and ({\bf 1, 35}) with $L = 1$, i. e. like
{\bf $\alpha \cdot \bigtriangledown$}.\\
\\
If we abstract relations from a pure quark picture without gradient
couplings, then this quark kinetic part of $\Theta_{\mu \nu}$ is all
there is apart from the trace contribution. In that case, we have the
equal time commutation relation for the whole energy operator:
\begin{displaymath}
\sum\limits_{r=1}^{3} \sum\limits_{i=1}^{8} \left[ {\cal F}_{ir}d^3 x,
\left[ \int {\cal F}_{ir} d^3 x, P_0 \right] \right] = 16/3 P_0 \, \, \, \,
{\rm + \, \, \, \, scale \, \, \, \, violating \, \, \, \, terms.}
\end{displaymath}
This relation, in the pure quark case, can be looked at in another way.
It is an equal time consequence of the relation
\begin{displaymath}
\Theta_{\mu \nu} = \lim_{y \rightarrow x} \frac{3 \pi^2}{32}
\partial_{\mu} \partial_{\nu} \left\{ \left( z^2 \right)^2
{\cal F}_{i \alpha} (x) {\cal F}_{i \alpha} (y) \right\} +
\, \, \, \, {\rm scale \, \, \, \, violating \, \, \, \, terms}
\end{displaymath}
that holds when the singlet tensor term in the light cone expansion of
${\cal F}_{i \mu} (x) {\cal F} _{j \nu} (y)$ is just proportional to
$\Theta_{\mu \nu}$ as in the pure quark case. This relatin is what, in
the pure quark version of the light cone algebra (extended to light cone
products), replaces the Sugawara$^{{\rm 23}}$ model, in which
$\Theta_{\mu \nu}$ is proportional to ${\cal F}_{i \mu} {\cal F}_{i\nu}$,
with dimension - 6. Our expression is much more civilized, having
$l = - 4$ as it should. A more general equal time commutator than the one
above, also implied by the pure quark case, is the following:
\begin{displaymath}
\sum\limits^{3}_{r=1} \left[ {\cal F} _{i r} (x), \partial_0 {\cal F}_{ir}
(y) \right] = 16i/3 \; \Theta_{00} \; \delta (x - y) \, \, \, \,
{\rm + \, \, \, \, scale \, \, \, \, breaking \, \, \, \, terms}.
\end{displaymath}

Another important point that should be emphasized is that the
$U_6 \times U_6$ algebra requires the inclusion of a ninth vector current
${\cal F}_{0 \alpha}$ and a ninth axial vector current 
${\cal F}^5_{0 \alpha}$, and that the Latin index for $SU_3$
representation components in Appendix II has to run from $0$
 to $8$. Now if the term in the energy density that breaks
 $SU_3 \times SU_3$ follows our usual conjecture and behaves like
 $- u_0 - cu_8$ with $c$ near $- \sqrt{2}$ and if the chiral symmetry
 preserving but scale breaking term $\delta $ is just a constant, then
 as $u \rightarrow 0$ scale invariance and chiral invariance become good,
 but the mass formula for the pseudoscalar mesons indicates that we do
 not want $\partial_{\alpha} {\cal F}_{0 \alpha}$ to be zero in that
 limit.$^{{\rm 10}}$ Yet ${\cal F}^5_{0 \alpha}$ is supposed to be
 conserved on the light cone. Does this raise a problem for the idea of
 $\delta =$ const. or does it really raise the whole question of the
 relation of the light cone limit and the formal limit $u \rightarrow 0$,
 $\delta \rightarrow 0$?\\
 \\
 If there are dilations, with $m^2 \rightarrow 0$ in the limit of scale
 invariance while other masses stay finite, how does that jibe with the
 light cone limit in which all masses act as if they go to zero?
 Presumably there is no contradiction here, but the situation should be
 explored further.\\
 \\
 Finally, let us recall that in the specific application of scaling to
 deep inelastic scattering, the functions $F(\xi)$ connect up with two
 important parts of particle physics. As $\xi \rightarrow 0$, if we can
 interchange this limit with the Bjorken limit, we are dealing with
 fixed $q^2$ and with $p \cdot q \rightarrow \infty $ and the behavior
 of the $F$'s comes directly from the Regge behavior of the corresponding
 exchanged channel. If $\alpha_p (0) = 1$, then $F^{ep}_2 (\xi) + F^{en}_2
 (\xi)$ goes like a constant at $\xi = 0$, i. e.,
 $\xi^{1- \alpha p (0)}$, while $F^{ep}_2 (\xi) - F^{en}_2 (\xi)$ goes
 like $^{1- \alpha \rho(0)}$, etc.\\
 \\
 As $\xi \rightarrow 1$, as emphasized by Drell and Yan$^{{\rm 8}}$,
 there seems to be a connection between the dependence of $F (\xi)$ on
 $1 - \xi$ and the dependence of the elastic form factors of the nucleons
 on $t$ at large $t$.\\
 \\
 \\
 {\bf 9. PROBLEMS OF LIGHT CONE CURRENT ALGEBRA}\\
 \\
 If we take the notion of current algebra on the light cone seriously
 we are faced with a number of important theoretical questions, to most
 of which we have already alluded. We shall attempt to summarize them
 here and to comment on them.\\
 \\
 We have exhibited in Eqs. (6.3) a closed algebraic system of light cone
 commutators of the connected parts of the 72 components of nine vector
 and nine axial vector bilocal currents, valid in the limit where all
 four points tend to lie on a straight line on the light cone (all six
 invariant intervals approachng zero). We shall refer to this system as
 the basic light cone algebra. The bilocal operators involved we may
 rename, in an obvious notation,
 ${\cal D} \left( x, y, \left( \frac{i \lambda_i}{2} \right) \gamma_{\mu}
 \right)$ and ${\cal D} \left( x, y, \left( \frac{i \lambda_i}{2} \right)
 \gamma_{\mu} \gamma_5 \right)$. They are well defined as $(x - y)^2
 \rightarrow 0$ and their local limits are ${\cal F}_{i \mu}(x)$ and
 ${\cal F}^5_{i \mu}(x)$ respectively. We may ask the following questions
 about the basic light cone algebra:
 \begin{enumerate}
 \item[a)] Assuming that further refinement of the SLAC experiments and
 work on corresponding neutrino experiments continue to support the basic
 algebra, what further practical experimental tests can be designed? We
 want to generalize the tests of light cone commutators of local currents
 to spin--flip matrix elements, to matrix elements with momentum transfer
 $\not= 0$, and to matrix elements between different numbers of particles.
 (We note, by the way, that as soon as we depart from matrix elements
 between 1 particle and 1 particle, the notion that mathematical
 dimension = physical dimension for the amplitudes is seen to be
 arbitrary, the mathematical dimension of the amplitude depending on the
 number of particles in a way that varies with our normalization. What
 must be preserved is the existence of a well--defined Bjorken limit for
 the commutator matrix elements, even though, with a given normalization
 convention, powers of masses occur in the final answer.)\\
 \\
 When many hadron momenta $p$ are present in the problem (all finite
 and timelike), we need a generalization of the Bjorken limit in
 momentum space, which corresponds to the light cone in co--ordinate
 space. Presumably, we choose a fixed light--vector $e$ and $a$ fixed
 timelike vector $a$ and write the current momentum $q$ as $ue + a$,
 where the variable $u$ is allowed to approach $\infty$. Then for any
 hadron momentum $p$, we have $2 q \cdot p \rightarrow 2 ue \cdot p$,
 while $q^2 \rightarrow 2 ue \cdot a$, and the ratios are all finite as
 $u \rightarrow \infty $ (since a timelike vector dotted into a
 light--like one is non--zero).
 \item[b)] Can tests be designed for the commutators of bilocal operators
 in the basic algebra, that is to say for light cone commutators of light
 cone commutators of currents?\\
 First, we should generalize the Bjorken limit further to cover more than
 one current momentum $q$. A possible way to do that may be to let
 $q_j = u_j e + a$, with fixed $e$ and $a$ as above. Then
 $q^2_j \rightarrow 2 u_j e \cdot a, \left( q_j + q_k \right)^2
 \rightarrow 2 \left( u_j + u_k \right) e \cdot a,\\
  2 q_j \cdot p_i
 \rightarrow 2 u_j e \cdot p_i$, etc. If all the $u$'s $\rightarrow \infty$,
 then there is a fixed ratio between any $q^2$ and the corresponding
 $2 q \cdot p_i$ in the limit.\\
 Next, we have to consider if we can really measure the light cone
 commutator of light cone commutators. Actually that is very difficult,
 and the tests may be practical only if we generalize, as discussed in i)
 below, from commutators of currents on the light cone to physical
 ordered products of currents as well.\\
 \\
 Tests of bilocal commutators are important not only for verifying that
 the bilocal algebra makes sense, but also because they involve the
 fourth powers of the quark charges, and therefore make possible comparison
 with the squares of the charges so as to check whether the fractional
 values are really correct. Other tests of the fractional charges are
 conceivable if the algebra is generalized to disconnected parts (hadron
 vacuum expectation values of commutators) as discussed in k) below, but
 there several questions arise that make a test within the basic algebra
 desirable.\\
 \item[c)] To what extent can we abstract the basic algebra from a quark
 field theory model with interactions? It is, of course, all right in a
 free quark model but so are a great many results that we would not dream
 of abstracting for real hadrons. Recent work of Llewellyn
 Smith,$^{{\rm 18}}$ Cornwall and Jackiw,$^{{\rm 24}}$ and Gross and
 Treiman$^{{\rm 25}}$ has confirmed that in a quark field theory model
 with neutral gluons, using formal manipulation of operators and not
 renormalized perturbation theory term by term, the basic algebra comes
 out all right in the presence of interactions. When the gluon is vector,
 the correspondence between the ${\cal D}$'s and quark expressions must
 be modified by the presence of the factor exp $i g \int^y_x B_{\mu} d
 l_{\mu}$, where $B_{\mu}$ is the gluon field, $g$ its coupling constant,
 and the integral is along a straight line.\\
 The renormalized perturbation theory, taken term by term, reveals various
 pathologies in commutators of currents. Not only are there in each order
 logarithmic singularities on the light cone, which destroy scaling, and
 violations of the rule that $\sigma_L / \sigma_T \rightarrow 0$ in the
 Bjorken limit, but also a careful perturbation theory treatment shows
 the existence of higher singularities on the light cone, multiplied
 by the gluon fields, such as we worried abuot earlier on the basis of
 dimensional analysis. For example,$^{{\rm 26}}$ in vector gluon
 theory we meet a term of the form
 \begin{displaymath}
 g (x - y)_{\alpha} \varepsilon_{\alpha \beta \gamma \delta}
 \partial_{\gamma} B_{\delta} / (x - y)^2
 \end{displaymath}
 
 occurring where we would expect from the basic algebra the finite operator
 ${\cal F}^3_0 (x, y)$: the gluon field strength, having in lowest order
 dimension $l = - 2$, can appear multiplied by a more singular function
 than can a finite operator ${\cal F}^5_0 (x, y)$ of dimension -3. Such
 a term would ruin the basic algebra as a closed system and even wreck
 the equal time algebra of charge densities by introducing a
 $\bigtriangledown \delta$ term), although leaving untouched particular
 commutators, such as those involved in the SLAC experiments, and in fact
 any matrix elements with $\Delta p = 0$. A term involving the gluon
 field strength would also elevate that operator to the level of a
 physical quantity, occurring in the light cone commutator of real
 local currents.\\
 If we wish to preserve the abstraction of the basic algebra, we must
 reject these ``anomalous'' singularities just as we do the logarithmic
 singularities in each order of renormalized perturbation theory and the
 occurrence of asymptotic longitudinal cross--sections. If, however, we
 blindly accept for hadrons the abstraction of any property of the
 gluon model that follows from naive manipulation of operators, we risk
 making some unwise generalizations of the basic algebra. It would be
 desirable to have some definite point of view about the relation of the
 abstracted results to the renormalized perturbatin theory. Such a point
 of view, if available, would replace the transverse momentum cut--off
 of Drell and collaborators as a way of forcing the barely renormalizable
 gluon models into the mold of a super--renormalizable theory.\\
 
 We note, for example, that if we take vector gluon theory at all
 seriously, we must deal with the fact that the vector baryon current
 ${\cal F}_{0 \mu}$ and the gluon exist in the same channel and are
 coupled, so that a string of vacuum polarization bubbles contributes to
 the unrenormalized current. But all the currents have fixed normalization,
 since their charges are well--defined quantum numbes, and it must be the
 unrenormalized currents that obey the algebra if the algebra is right.
 Hence, if the renormalized coupling constant $g_1$, of the gluon is to
 be non--zero, its renormalization constant $Z_3^{-1}$ must be finite
 and we must imagine that the sum of perturbation theory yields the
 special case of a ``finite vector theory''$^{{\rm 27}}$, if we are
 to bring the vector gluon theory and the basic algebra into harmony.
 Perhaps this picture of a ``finite theory'' (assuming it is consistent,
 and we note that it involves finding roots of a particular equation
 for the coupling constant, an equation which may not have roots!) leads,
 when the perturbation theory is summed, to canonical scaling and the
 disappearance of the ``anomalous'' light cone singularities, so that
 the basic algebra is preserved. But, if that is so and we lean on the
 ``finite theory'' for our abstraction of the algebra, we have trouble
 with the possible generalization of the algebra to disconnected parts with
 the accompanying naive or free quark behaviour at high momentum of the
 vacuum expectation values of bilocal operators
 [as discussed in $k)$ below]. The reason is that in the ``finite
 theory'' the asymptotic behavior of the vacuum expectation value of
 current products or current commutators has a reduced singularity compared
 to naive or free quark behaviour; this is evident in the case of two baryon
 currents in order to make $Z_3^{-1}$ finite. Thus the logic of the
 ``finite theory'', while it might preserve the basic algebra, excludes the
 simplest generalization to disconnected parts and may exclude other
 generalizations.\\
 \item[d)] Is a generalization possible to a connected light cone algebra
 of 144 components of $V, A, S, T,$ and $P$ densities as in the free quark
 model, with divergences of the vector and axial vector currents given by
 $S$ and $P$ densities with definite coefficients (effective quark masses)
 and with divergence and curl of the tensor current given by well--defined
 quantities in the theory?\\
 Using formal manipulation of operators, all of this seems to happen in the
 quark theory model with vector gluons. (If the gluons are scalar or
 pseudoscalar, the various divergencies do not come out in terms of
 densities in the algebra.)\\
 The resulting generalized system has densities
 \begin{displaymath}
 {\cal D} \left( x, y, \frac{i \lambda_i}{2} \gamma_5 \right); {\cal D}
 \left( x, y, \frac{\lambda_i}{2} \right) \, \, \, {\rm and} \, \, \,
 {\cal D} \left( x, y, \frac{\lambda_i}{2}\sigma_{\mu \nu} \right)
 \end{displaymath}
 
 as well as the vector and axial vector ones of the basic algebra, and
 the system closes algebraically under commutation, with the same rules
 as the free quark theory. Besides the familiar divergence equations
 \setcounter{page}{1}
\setcounter{equation}{0}
\renewcommand{\theequation}{9.\arabic{equation}}
 \begin{eqnarray}
 \frac{\partial}{\partial x_{\mu}} \left( x, x, \frac{i \lambda_i}{2}
 \gamma_{\mu} \right) & = & {\cal D}
 \left( x, x, \frac{i\left[ M, \lambda_i \right]}{2}  \right) \, ,
  \\ \nonumber \\
 \frac{\partial}{\partial x_{\mu}} {\cal D} \left( x, x,
 \frac{i \lambda_i}{2} \gamma_{\mu} \gamma_5 \right) & = & {\cal D}
 \left( x, x, \frac{i \left\{ M, \lambda_i \right\} \gamma_5}{2} \right)
 \, ,
 \end{eqnarray}
 
 where $M$ is the ``quark mass'' matrix, diagonal for the three quarks
 $u, d,$ and $s,$ we have in addition relations for the tensor currents:
 
 \begin{eqnarray}
 \frac{\partial}{\partial x_{\mu}} & {\cal D} & \hspace*{-0.3cm}
 \left( x, x, \frac{\lambda_i}{2} \sigma_{\mu \nu} \right) \, ,
 \\
 \nonumber \\
 & = & - {\cal D}
 \left( x, x, i \left\{ M, \lambda_i/2 \right\} \gamma_{\nu} \right) +
 \left[ \left( \frac{\partial}{\partial x_{\nu}}
 \frac{\partial}{\partial \gamma_{\nu}} \right)
 {\cal D} \left( x, y, i \lambda_i/2  \right)  \right]_{x=y}
 \nonumber \\
 \nonumber \\
 \frac{\partial}{\partial x_{\mu}} & {\cal D} & \hspace*{-0.3cm}
 \left( x, x, \frac{ \lambda_i}{2} \sigma_{\mu \nu} \gamma_5 \right)
 \\
 \nonumber \\
 & = & - {\cal D} \left( x, x, \frac{i \left[ M, \lambda_i \right]}{2}
 \gamma_{\nu} \gamma_5 \right) + \left[ \left(
 \frac{\partial}{\partial x_{\nu}}
 \frac{\partial}{\partial y_{\nu}} \right) {\cal D}
 \left( x, y, \frac{i \lambda_i}{2} \gamma_5 \right) \right]_{x=y^.}
 \nonumber
 \end{eqnarray}
 In the first of these we see the generalization of the famous Gordon
 break--up of the Dirac vector current into a ``convective current''
 and the divergence of a tensor ``spin current''. In the second, we see
 appearing on the right--hand side the axial vector analogue of the
 ``convective current'' and we note that it is a ``second--class current''
 that may some day play a r\^ole in a theory of CP violation. It is
 fascinating that these convective currents occur in the generalized
 algebra as first internal derivatives of the bilocal quantities.\\
 It is interesting to look into the divergences not only of the local
 currents but also of their internal derivatives. In a free quark model
 the bilocal vector currents corresponding to conserved local vector
 currents are themselves conserved (with respect to $x + y$); in other
 words, all their internal derivatives are conserved. This is an example
 of an outrageously strong result that we presumably do not wish to
 abstract, and indeed it fails in a quark model with interactions.\\
 Let us look in detail at the divergence of the first internal derivative
 of the baryon current in the vector gluon model, putting $R = (x + y)/2$
 and $z = x - y$. These first internal derivatives are light cone
 quantities and defined in the basic algebra. The equations of motion
 yield:
 \begin{equation}
 \frac{\partial}{\partial R_{\mu}} 
 \left[ \frac{\partial}{\partial z_{\nu}}
 {\cal D} \left( R, z, i \gamma_{\mu} \right) \right]_{z = 0}
 = - ig \left( \frac{\partial B_{\mu}}{\partial R_{\nu}} -
 \frac{\partial B_{\nu}}{\partial R_{\mu}} \right) {\cal D}
 \left( R, 0, i \gamma_{\mu} \right) \, .
 \end{equation}
 
 We may also look at the first nonlocal correction to Eq. (9.2):
 \begin{eqnarray}
 \frac{\partial}{\partial R_{\mu}} {\cal D} \left( R, z,
 \frac{i \lambda_i}{2} \gamma_{\mu} \gamma_5 \right) & = & {\cal D}
 \left( R, z, i \frac{i \left\{ M, \lambda_i \right\}}{2} \gamma_5 \right)
 \nonumber \\
 \\
 & & - igz_{\nu} \left( \frac{\partial B_{\mu}}{\partial R_{\nu}} -
 \frac{\partial B_{\nu}}{\partial R_{\mu}} \right) {\cal D} \left( R, z,
 \frac{i \lambda_i}{2} \gamma_{\mu} \gamma_5 \right)
 \nonumber \\
 \nonumber \\
 & & + \cdots \cdots \cdots \cdots  \nonumber
 \end{eqnarray}
 
 The first of these relations shows how, in the vector, gluon model the
 integral of the first internal derivative of the baryon current fails to
 be conserved and is therefore not equal to the total momentum that
 corresponds to the failure of the sum rule (7.16), which is now being
 tested by neutrino experiments. It will be exciting to see whether
 experiment leaves room for gluons or not in our abstraction of algebraic
 results from models.\\
 The second relation is important in a different way, since it shows how,
 if an anomalous linear singularity on the light cone is introduced into
 ${\cal D} \left( R, 0, i\gamma_{\mu} \gamma_5 \right) \propto
 {\cal F}^5_{0 \mu} (R)$, an ``anomalous divergence term'' $\propto g^2
 \varepsilon_{\mu \nu k \lambda} \times B_{\mu \nu} B_{\kappa \lambda}$
 appears
 to be introduced into the divergence of the ninth axial vector current.
 This anomalous divergence has dimension -4 (which we supposed could not
 be present in $\partial_{\mu} {\cal F}^5_{i \mu}$) and spoils the
 situation in which the divergences or currents are contained in the
 generalized algebra.\\
 It is unclear whether the mathematical relation between the high energy
 anomalous singularity and the low energy anomalous divergence is real or
 apparent, when operator products are carefully handled. Like the Adler
 term discussed in j) below, the anomalous divergence may be obtainable
 as a kind of low energy theorem and might survive a treatment in which
 the anomalous singularity is gotten rid of.
 \end{enumerate}
 Summarizing what we have just examined, we add two more questions to
 our list:
 \begin{enumerate}
 \item[e)] Is there a failure of the sum rule (7.17) and thus room in
 the algebraic structure for abstraction from a model with gluons?\\
 Na\"ive manipulation of operators in the vector gluon model seems to
 give the enlarged light cone algebra of the connected parts of 144
 densities, with no anomalies singularities and no anomalous
 divergences.\\
 The renormalized perturbation theory, taken term by term, contains a
 large number of anomalous that spoil even the basic algebra, though not
 necessarily in direct conflict with experimental results so far.\\
 A ``finite theory'' approach is needed if the sum of renormalized
 perturbation theory is to be brought into any sort of correspondence
 with the light cone algebra. However, it is not at all clear how many
 anomalies are cured in this manner. (We note, by the way, that for the
 scalar and pseudoscalar densities to have canonical dimensions and for
 their unrenormalized versions to be finite, so that they can obey the
 algebra and allow finite bare quark masses $M$ as coefficients,
 {\it another} function of the coupling constant must vanish, namely, the
 exponent that appears in the mass renormalization in the finite theory.)\\
 In any case, certain anomalies, like the anomalous divergence of
 ${\cal F}^5_{o \kappa}$, come out only in lowest order of renormalized
 perturbation theory and appear difficult if not impossible to fix by the
 ``finite theory'' approach, even if other diseases are cured.\\
 We are left, then, with four possible attitudes:
 \begin{enumerate}
 \item[A)] The whole system, including scaling, is wrong as in
 renormalized perturbation theory term by term.
 \item[B)] A ``finite theory'' approach is to be used, from which certain
 features of canonical scaling can be abstracted, but in which a number
 of anomalies are left that wreck either the basic or the enlarged algebra
 as a closed system, while also destroying the possibility of abstracting
 the behaviour of disconnectd parts from free quark theory or naive
 considerations.
 \item[C)] The na\"ive approach is right, and the basic algebra can be
 abstracted, with probably the enlarged algebra as well, and perhaps even
 the behaviour of disconnectd parts. The gluon field is not necessarily
 directly observable, but its effect is felt indirectly, for example,
 i. e., the failure of the sum rule (7.16). In this case, what happens to
 the Adler anomaly, discussed in j) below, which formally resembles the
 corresponding gluon anomalies, but involves the real electromagnetic
 field and real electric charges, instread of the presumably fictitious
 gluon quantities?
 \item[D)] The na\"ive approach is right, but we are forced to have the sum
 rule (7.16) and the corresponding conservation of the quark momentum
 alone, as if we had a super--renormalizable self--interaction of quark
 currents. This last situation seems attractive, as we have indicated in
 earlier Sections, but is it right? What will experiments have to say about
 it? Is it consistent theoretically?\\
 \end{enumerate}
 \item[f)] Are there anomalous divergences in hadron theory or do these
 go away if the anomalous singularities disappear?\\
 We might remarkt, by the way, that an anomalous divergence term in
 ${\cal F}_{0 \mu} \, ^5$ looks at first sight like a welcome addition,
 since it distinguishes the ninth axial vector current from the other
 eight and seems to provide an excuse for the apparent failure of the
 ninth one to have a zero mass pseudoscalar meson in the approximation
 in which $M$ is neglected, while the other eight have the pseudoscalar
 octet; the distribution of mass squared of the nine pseudoscalar mesons
 certainly suggests some sort of distrinction. However, in fact an
 anomalous divergence is not needed to provide such an excuse, since the
 algebra $U_3 \times U_3$ of the vector and axial vector current charges
 already allows for a distinction. Since the charge $F_0^5$ commutes with
 all the others, there is no reason for it not to vanish when $M$ is
 neglected, unlike the other eight, which are prevented from vanishing
 by their commutation rules. Thus $F_0^5$ escapes the choice, in the
 approximation of its conservation, between having a zero mass pseudoscalar
 meson and causing degeneracy of opposite parities, while the other
 $F_i^5$ do not escape the choice and apparently have the massless
 pseudoscalar octet in the limit $M \rightarrow 0$.\\
 \item[g)] Is there any practical way of testing the enlarged algebra
 inclusing the divergences of vector and axial vector currents?\\
 We have alluded to this matter in previous sections when we discussed
 tests of the dimensionalities of these divergences, which are here equal
 to -3. Weak interaction tests are perfectly possible, but they are very
 difficult, especially since the amplitudes of leptonic processes involving
 the current divergences vanish with the lepton masses.
 \item[h)] Assuming the extended algebra, are we right in our understanding
 of the relation between high energy pion elastic scattering and the
 Bjorken limit of the matrix element of the commutator of two
 pseudoscalar densities?\\
 We commute a pseudoscalar density, say
${\cal D} \left( x, x, \left( \left( i \lambda_3/2 \right) \gamma_5
\right) \right)$, with itself at two points near the light cone and obtain
at the righthand side a term proportional to $d_{33j} \partial_{\mu}
\left[ \varepsilon \left( x_0 - y_0 \right) \delta \left( \left(
x - y \right)^2 \right) \right]$. ${\cal D} \left( x, y,
\left( i \lambda_j /2
\right) \gamma_{\mu}  \right)$, which, between two proton states of
equal momenta, gives just the SLAC form factors and the related one for
neutrino experiments, provided we take the Bjorken limit. We then
utilize the principle invoked in Section 8 that we may interchange the
Bjorken limit and the limit $\xi \rightarrow 0$ to obtain the high
energy limit for fixed large $q^2$ and the connection with Regge behaviour.
Assuming, as before, that $\alpha_p(0) = 1$, we get a form factor
$\sim \xi^{-1}$ at small $\xi$, compatible with the SLAC results, and the
amplitude for the commutator at high energy for fixed large $q^2$ goes
like $(2 p \cdot q) / q^2$, or $s^1/q^2$.\\
Now, for any $q^2$, we expect this matrix element of the commutator of
pseudoscalar densities to go like $s^1 \varphi \left( q^2 \right)$, since
$a_{\rho} (0) = 1$. At $q^2 = - m^2_{\pi}$, $\varphi $ should have a double
pole corresponding to the pion scattering. If PCAC is useful here, then the
double pole should dominate the behaviour of $\varphi $ near $q^2 = 0$ and
in that region we can calculate $\varphi $ from the asymptotic elastic pion
scattering amplitude (i. e., the total cross--section), the Goldberger--Treiman
constant, and the ``quark masses'' in the diagonal matrix $M$, which
have a definite physical significance in the extended algebra, since
they relate the divergences of the axial vector currents to the
pseudoscalar densities in the algebra.\\
Thus we know the behaviouf of $\phi \left( q^2 \right)$ at large
$q^2$ (proportionality to $1/q^2$ with a coefficient obtainable from the
usual deep inelastic form factors) and we know it at small $q^2$ in terms
of the ``quark masses'' and the total asymptotic pion cross--section.
Unfortunately we do not know any reliable way to connect the two regions,
but some day this insight may be helpful. Anyway, we see that the extended
algebra is perfectly compatible with Regge behaviour and the interchange of
limits.\\
\item[i)] To what extent can we generalize the algebra to a set of
relations for the connected parts of products, or of physical ordered
products, of operators near the light cone?\\
First of all, if the commutator algebra is correct, the generalization
to ordinary binary products near the light cone seems straightforward; we
need only exclude catastrophic singularities in the anticommutator (or
real part in momentum space) near $z = 0$. Then the operator product near
the light cone looks like the commutator, but with $\varepsilon
\left( z_0 \right) \delta \left( z^2 \right)$ replaced by $\left( 2 \pi i
\right)^{-1} \left( z^2 - iz_0 \varepsilon \right)^{-1}$.\\
Next, we go on to binary ordered products of currents, as in Feynman
amplitudes. To clarify the ideas, let us look at the ordered product of
two electromagnetic currents and see what would happen if we were
abstrac\-ting our formula from a model containing a scalar charged field
$\phi $. Then there would be a non--vanishing asymptotic longitudinal
cross--section $\sigma_L$, and terms in $\delta_{\mu \nu} - \left( q_{\mu}
q_{\nu} \right) / q^2$ would survive in the Bjorken limit. Multiplying
such terms in co--ordinate space by $\varepsilon \left( z_0 \right)$
would be non--covariant, and it would be necessary to add non--covariant
terms to the ordered product to restore the covariance; these correspond
to operator Schwinger terms and they would be proportional to $\Phi^+
\Phi$. In addition, to make up the physical two--photon amplitude, it would
be necessary to add in the seß-cond order physical ``sea--gull''
interaction $e^2 A_{\mu} A_{\mu} \Phi^+ \Phi$, a covariant term of second
degree in the electromagnetic potentials and also propotional to
$\Phi^+ \Phi$.\\
The essential point to be learned from this example is that it is only
the complete amplitude, or {\it physical} ordered product, including all the
possible types of contribution mentioned above, that matters. For
electromagnetism, that representes the actual coupling to two virtual
photons to order $e^2$.\\
Given our picture of the ordinary product of two electromagnetic currents
near the light cone, is it trivial to construct the physical ordered
product, just replacing $\left( z^2 - iz_0 \varepsilon \right)^{-1}$ by
$\left( z^2 - i \varepsilon \right)^{-1}$? We can presumably dispense with
the complications just mentioned for the abstraction from charged scalar
theory, since we have no asymptotic $\sigma _L$, no operator Schwinger
terms, and presumably no explicit ``sea--gulls'' of the type encountered
there. However, we must be careful about the possibility of some subtler
type of subtraction term in the dispersion relation connecting absorptive
and dispersive parts of the physical amplitude. Further investigation of
that point would be very useful, and should soon clear up the matter.\\
If the connected part of the physical ordered product of two
electromagnetic currents is simply understood as we have indicated, then
we are in a position to propose experimental tests of the bilocal algebra
by experiment. For example, we examine the reaction $e^- + p \rightarrow
e^- + X + \mu^+ \mu^-$, where $X$ is any hadronic system, and consider
the cross--section summed over $X$, which gives us the amplitude of a
fourth order electromagnetic process, with a proton as initial and final
state and $\Delta p = 0$. There are two variables $q$, one for the
electrons and one for the muon, and we go to the generalized Bjorken
limits, as sketched above. We are dealing with the light cone commutator
of two light cone--physical ordered products, and if the latter are indeed
simple, we have the light cone commutator of two bilocal currents, with
all four points tending to lie on a straight line. The right--hand side
then involves the same matrix elements of bilocal currents as in the SLAC
and corresponding neutrino experiments, and a test of bilocal algebra and
of quark charges becomes possible in principle, as suggested above under
b).\\
Theoretical investigation should be extended to the physical ordered
light cone product of any number of electromagnetic currents, to see if
surprises turn up.\\
Finally, let us allude to the generalization from electromagnetic currents
to others in the system, when we take physical ordered products. Except
for PCAC considerations, as mentioned below under j), we can attach
meaning to physical ordered current products only if we discuss the actual
physical interactions to which they refer. In other words, we must
consider products of weak currents or mixed products of weak and
electromagnetic currents and ask about the actual amplitudes for weak
processes or weak electromagnetic processes, to the lowest order in $G$
and $e$ in which these occur. (Indeed, if $G$ is really like
$e^2 m_X^{-2}$, where $m_X$ is an intermediate boson mass, then we may
have to treat weak and electromagnetic orders as interchangeable.) Such
discussions contain considerable uncertainties, since the amplitudes may
contain intermediate boson propagators, electromagnetic vertices of
intermediate bosons, and more complicatd Yang--Mills type interactions of
intermediate bosons. We would have to base our work on a definite picture
of higher-order weak and weak--electromagnetic processes in order to make
it fully meaningful and understand the significance of any subtraction
terms that arise. The same statement may be turned arount, however, to
sound more hopeful: a study of the physical ordered products near the
light cone of weak and electromagnetic currents can help in the
construction of a skeleton theory of higher order weak and
weak--electromagnetic processes.\\
\item[j)] What are the implications for light cone current algebra of the
``Adler anomaly''?\\
Here we must turn our attention to the physical amplitude for two photons
to turn into the divergence of the axial vector current
${\cal F}^5_{3\alpha}$, where the physical significance of the last is
given not by the weak interaction but the PCAC hypthesis, treating the
pion mass as small and obtaining an approximation to the decay amplitude
$\pi^0 \rightarrow 2 \gamma$. The photon frequencies may be treated as
small, also, and the whole problem can be phrased as a search for a low
energy limit.\\
In the renormalized perturbation theory approach to the vector gluon model,
a sophisticated treatment shows that Adler's ``anomalous divergence'' term
in $\partial_{\alpha} {\cal F}^5_{3 \alpha}$, of the form (const.)
$e^2 F_{\mu \nu} F_{\mu \nu}^*$, shows up only in the zeroth order of the
renormalized perturbation expansion in $g^2$, and thus the PCAC
approximation to $\pi _0 \rightarrow 2 \gamma $ can be calculated exactly
in terms of a simple triangular quark loop, which gives the value of the
constant.\\
If we look at the Adler calculation in terms of a vacuum closed loop it
seems to belong with our discussion under k) of disconnected parts, but if
we think of it as concerning the matrix element between vacuum and a low
mass $\pi^0$ of the physical ordered product of two electromagnetic
currents, it is seen to be related to a connected part. Again, the Adler
result is a ``low energy theorem'', but it is connected with a possible
high energy singularity arising through electromagnetic effects, in a
way that parallels the apparent relation between anomalous divergence and
high energy singularity discussed for the
vector gluon model, with the difference noted above that photons are real
and gluons presumably fictitious.\\
Now the actual calculation of the $\pi^0 \rightarrow 2 \gamma $ decay
amplitude by the Adler method gives, for quarks with Fermi--Dirac
statistics, an amplitude about three times too small to agree with
observation, while ``parastatistics of rank three'' gives a factor of
three and good agreement with experiment.\\
We note that when using quarks as constituents of hadrons in the simple
phenomenological $3q$ picture of the baryon, those ``constituent quarks''
look as if they should be assigned para--Fermi statistics of rank three,
so that we can have for the ground state a totally symmetric rather than
a totally antisymmetric spatial wave function, the latter being rather
bizarre. We discuss below under m) the complicated transformation
connecting these ``constituent quarks'' (non--relativistic for a hadron
at rest and with low probability for pairs) with the relativistic
``current quarks'' of the quark--gluon field theory model (in which a
hadron bristles with $q \bar q$ pairs). That transformation presumably
does not affect the statistics. Thus, including the Adler result, we have
an argument in each case for parastatistics.\\
We may describe the ``paraquarks'' in the following way. We start with
three kinds of $s$ quark, three kinds of $u$ quark, and three kinds of
$d$ quark, nine in all, obeying conventional Fermi--Dirac statistics, and
then apply a supplementary condition that any physical hadron system is
in a singlet state of the new $SU_3$ spin. This supplementary condition
is presumably not allowed if the quarks are real, since it does not
factor when a system is divided into two distant subsystems. Thus we are
dealing with three fictitious ``paraquarks'' $u, d$, and $s$.\\
If we insist on having real quarks, then the Adler argument leads us to
nine real particles, giving us a so-called ``three--triplet''
situation.\\
The paraquarks always give us a factor of 3 in a vacuum loop compared to
Fermi--Dirac quarks. This is important when we go on to our next
generalization, which is to disconnected parts of amplitudes.\\
\item[k)] Apart from the ``Adler anomaly'', to what extent can we use
here quark theory on the light cone for the algebra of disconnected
parts of the currents and for the vacuum expectation values of bilocal
currents?\\
We have mentioned briefly in Section 6 the possibility that free quark
behaviour might characterize not only the algebraic structure of
connected amplituds but also the high frequency limits of expectation
values of current producs or commutators in the hadron vacuum. We still
do not know whether that makes sense or not, and whether, if it makes
sense, it is experimentally correct.\\
We mentioned the simplest consequence of using abstracting light cone
formulae between vacuum and vacuum, namely the prediction of the
asymptotic total cross--section to order $e^4$ for $e^+ + e^- \rightarrow $
hadrons divided by the same for $e^+ + e^- \rightarrow \mu^+ + \mu^-$.
With Fermi--Dirac quarks, we would get an asymptotic ratio of
$\left( 2/3 \right)^2 + \left( - 1/3 \right)^2 + \left( - 1/3 \right)^2
= 2/3$, but with paraquarks we get three times as much, namely 2. The
explicit check on the fractional charges in the model is, of course,
less convincing now that one predicts 2, and the r\^ole of experimental
tests depending on connecting parts becomes more important.\\
A fourth order test of the disconnected commutator of ordered light cone
products is provided by the cross--section for
$e^+ + e^- \rightarrow \mu^+ + \mu^- + X$, summed over $X$, as suggested
by Gross and Treiman.$^{{\rm 25}}$ Again the result should be multiplied
by 3 for parastatistics.\\
We must consider here the possibility that the vacuum expectation values
of current products are less singular at high frequencies than in the
free quark model; such a situation obtains, for example, in the
``finite theory'' approach to the vector gluon model. In such a case,
the cross--section ratio
\begin{displaymath}
\sigma \left( e^+ + e^- \rightarrow \, \, \, {\rm hadrons} \, \, \, 
\right) / \left( \sigma \left( e^+ + e^- \rightarrow \mu^+ + \mu^- \right)
\right)
\end{displaymath}
would tend asymptotically to zero in lowest order in electromagnetism,
instead of 2/3 or 2. These considerations make the possible experimental
investigation of the high energy behaviour $\sigma \left( e^+ + e^-
\rightarrow \, \, {\rm hadrons} \, \, \right)$ expecially interesting.
Unfortunately, the energy of colliding beam expe\-riments now envisaged is
limited to a total of 7 GeV. Furthermore, there is a practical limitation
at sufficiently high energy, when higher order electromagnetic effects
make the one--photon annihilation difficult to measure.\\
We note that the high energy behaviour of the vacuum expectation value
of the product or commutator of two scalar or pseudoscalar densities is
important, as well as that of two vector or axial vector currents. Much
recent theoretical work on the $K_{e3}$ and $K_{\mu 3}$ decays has been
based on the notion that the Fourier transform of the vacuum expectation
value of the ordered product of two such densities obeys an unsubtracted
dispersion relation, whereas free quark theory would suggest two
subtractions. Where does the truth lie?\\
\item[l)] If we assume the basic bilocal algebra, or go further and assume
some of the generalizations discussed here, do we at some point abstract
so much from a quark model that we end up with the necessity of having real
quarks (or three real triplets)?\\
We have alluded to this all--important question in Section 8. It is still
not cleared up. If the bilocal algebra is to be maintained without real
triplets, we must somehow benefit from what is effectively the asymptotic
form $\left( i \gamma \cdot p \right)^{-1}$ of free quark propagators
without having any actual propagation of particles with single quark
quantum numbers; instead singularities occur only for mesons and baryons,
etc., with quark number divisible by 3. No one knows how to write down
explicitly a field theory in which the quarks are permanently bound and
nevertheless act free at large momenta, but that sort of thing is what
we seem to require of the abstract hadron theory.\\
Meanwhile, it would be useful to investigate further whether any of the
assumptions discussed here can be shown to lead to real triplets.\\
\item[m)] Do the considerations discussed here throw any light on the
nature of the transformation connecting ``constituent quarks'' and
``current quarks''?\\
Let us first put the question into a more physical form, quarks being after
all probably fictitious entities. We note that the constituent quark
model has a rough symmetry under a group $SU_6 \times SU_6 \times O_3$,
respresenting, respectively, the spin and unitary spin
of quarks, and the relative orbital angular momentum. For collinear
processes (with all particles moving in the $z$ direction, say) the
approximate symmetry reduces to the famous subgroup
$\left( SU_6 \right)_W \times O_2$. We may examine the special case of
$P_z = \infty$ and the resulting $\left( SU_6 \right)_W$ group
$\left( SU_6 \right)_{W- \infty, strong}$.\\
Now by studying the charges associated with various currents at
$P_z = \infty$ (looking at matrix elements between finite mass states),
we arrive at the algebra of another $\left( SU_6 \right)_W$, which we
may call $\left( SU_6 \right)_{W, \infty, currents}$. Between finite mass
states at $P_z = \infty$ we have
\begin{eqnarray}
F_i & = & \int {\cal D} \left( x, x, \frac{\beta \lambda_i}{2} \right) \,
d^3 x = \int {\cal D} \left( x, x, \frac{\beta \lambda_i}{2} \alpha_z
\right) \, d^3 x, \nonumber \\ \nonumber \\
- F_i^5 & = & \int {\cal D} \left( x, x, \frac{\beta \lambda_i}{2}
\sigma_z \right) \, d^3 x = - \int {\cal D}
\left( x, x, \frac{\beta \lambda_i}{2}
\gamma_5 \right) \, d^3 x, \nonumber \\  \\
F_{ix} & \equiv & \int {\cal D} \left( x, x, \frac{\beta \lambda_i}{2}
\beta \sigma_x \right) d^3 x \, = - \int {\cal D} \left( x, x,
\frac{\beta \lambda_i}{2} i \beta \alpha_y \right) \, d^3 x,
\nonumber \\ \nonumber \\
F_{iy} & \equiv & \int {\cal D} \left( x, x, \frac{\beta \lambda_i}{2}
\beta \sigma_y \right) d^3 x \, = \int {\cal D}
\left( x, x, \frac{\beta \lambda_i}{2}
i \beta \alpha_x \right) \, d^3 x \, . \nonumber
\end{eqnarray}

[If we do not wish to include tensor currents, we may discuss instead
just the subgroups
\begin{displaymath}
\left( Su_3 \times SU_3 \right)_{W,\infty, strong} \, \, {\rm and} \, \,
\left( SU_3 \times SU_3 \right)_{W, \infty, currents} \, .
\end{displaymath}

[If we do not like to work at $P_z = \infty$, we can switch to the
construction of light--like charges and construct an
$\left( SU_6 \right)_{W, currents}$ or an
$\left( SU_3 \times SU_3 \right)_{W, currents}$ out of those.]\\
What is the relation between $\left( SU_6 \right)_{W, \infty, strong}$
and $\left( SU_6 \right)_{W, \infty, currents}$? That is a physical
question that can replace the question about the relation of constituent
quarks to current quarks. There must be a transformation, perhaps a unitary
transformation, taking the generators of one $\left( SU_6 \right)_W$ into
those of the other. We know that this transformation is very diffe\-rent
from
the unit transformation, since $\left( SU_6 \right)_{W, \infty, strong}$
is approximately conserved, while
$\left( SU_6 \right)_{W, \infty, currents}$ is very far from conserved. We
know that baryon and meson eigenstates of mass are very impure with
respect to $\left( SU_6 \right)_{W, \infty, currents}$. If they were pure,
there would be no anomalous magnetic moments for neutron and proton,
$- G_A / G_V$ would be 5/3, etc. Furthermore, we can see from many
arguments, for example the one about anomalous moments, that the
transformation between the two $\left( SU_6 \right)_{W}$'s mixes up
orbital angular momenta. It mixes {\bf 56}, $L = 0^+$ with {\bf 70},
$L = 1^-$, for example, at $P_z = \infty$. In ``first appriximation'',
so to speak, the correction to unity in the transformation behaves like
{\bf 35}, $L = 1$ under either $\left( SU_6 \right)_W$. We note, to avoid
confusion, that the charges $F_i$ are not much affected by the
transformation, and $I$ and $Y$ not at all.\\
Now a hint about the transformation is provided by PCAC. (We are indebted
to Mr. H. J. Melosh and Mr. J. Amarante for discussions of this point.)
Assuming the generalized algebra of 144 densities, we use PCAC to tell us
that at low frequencies the pseudoscalar densities act like fields for
pseudoscalar mesons. Now under commutation with appropriate generators
of $\left( SU_6 \right)_{W, \infty, currents}$, the pseudoscalar densities
are transformed into components of vector currents ${\cal F}_{l \mu}$.
But under commutation with analogous generators of
$\left( SU_6 \right)_{W, \infty, strong}$, the effective ``pseudoscalar
meson fields'' are transformed into components of effective ``vector meson
fields''. In order for the transformation to be different from unity, the
effective ``vector meson fields'' must be different from the regular
vector currents ${\cal F}_{i \mu}$. Suppose the transformation is
unitary; then if we make an expansion of it about 1 as $U = 1 + i A +
\cdots$, we can discuss some properties of $A$. A possible expansion would
involve dimensionality. The regular vector current would go into itself,
with $l = - 3$, plus a correction with $l = - 4$, plus another correction
with $l = - 5$, etc. In a model where effective ``bare quark masses''
$M$ have real physical meaning (relating divergences of vector and axial
vector currents to scalar and pseudoscalar densities of the algebra), we
might think of the expansion as one in inverse powers of those masses,
even though the masses are probably small and the expansion very bad in
practice: we may still learn somethng from it.\\
We have, then, $A$ as an operator that commutes with the pseudoscalar
densities
${\cal D} \left( x, x, \left( i \lambda_i / 2 \right) \gamma _5 \right)$
to give correction with $l = - 4$ in the effective ``vector meson field''
to the vector currents
${\cal D} \left( x, x, \left( i \lambda_i / 2 \right) \gamma_{\mu} \right)$.
What can these corrections be? The logical candidates are the convective
currents $\left[\partial / \partial z_{\mu} {\cal D} \left( R, z,
\left( i \lambda_i / 2 \right) \right) \right]_{z=0}$, where
$R = \left( x + y \right) / 2$ and $z = x - y$.These have just the right
properties. The operator $A$ behaves like {\bf 35}, $L = 1$, has the
right charge conjugation behaviour, and so forth.\\
If we look explicitly at the vector gluon model, we see that the first
order transformation $1 + iA$ corresponds to the first order expansion
of the Foldy--Wouthuysen transformation, and this may be an important clue
to the nature of the whole transformation that connects constituent
quarks and current quarks.\\
Confusion has existed for many years between the two kinds of quarks, and
study of the transformation may help to clear up such confusion. Many
theorists have been surprised to find that the current quarks (or
``partons'') in the deep inelastic scattering analysis of the proton show
indefinitely large numbers of pairs, while the constituent quarks in the
quark model of the proton are three in number, with little allowance
for pairs. The fact that the transformation is very far from unity, of
course, explains the difference.\\
\item[n)] What can we do to explore the connection, if any, between scaling
in high energy hadronic processes and scaling in electromagnetic and
weak processes?\\
High energy hadronic scaling has been interpreted by Mueller$^{{\rm 28}}$
as coming from the applicability of Regge theory to many particle
processes at high energy, with the leading Regge exchange being that of
a Pomeranchuk pole with $a_p (0) = 1$. (It is not yet certain whether
this pole has to be a moving one; there is perhaps still a possiblity
that it might be a fixed pole and the Gribov--Pomeranchuk difficulty
overcome by the existence of a moving singularity that passes $a = 1$
between $t = 0$ and the lowest hadron threshold $t = 4 m^2_{\pi}$ for the
$P$ channel.) Mueller then gets, for an inclusive reaction with incoming
momenta $p$ and $p'$ and an outgoing momentum $q$ for the particle
observed forward scaling (say) when $q \cdot p$ and $p \cdot p'$ go to
$\infty $ proportionately with $q \cdot p$ finite, backward scaling when
$q \cdot p'$ and
$p \cdot p'$ go to $\infty $ proportionately with $q \cdot p$ finite, and
``pionization'' when $q \cdot p$ and $q \cdot p'$ both go to infinity
like $\sqrt{p \cdot p'}$. Here $p^2, p'^2$, and $q^2$ are, of course,
fixed.\\
No one seems really to understand the connection between this kind of
scaling and the light cone scaling for weak and electromagnetic currents
that we have discussed, with the corresponding Bjorken limits in which
various quantities $q^2 \rightarrow \infty $ for current momenta $q$.
This is true despite the fact that by interchanging limits one can
relate light cone scaling as $\xi \rightarrow 0$ Regge behaviour for
large $q^2$.\\
One common feature of hadronic and light cone scaling is the effective
transverse momentum cut--off in both cases, and that may provide a clue
to a possible connection when we understand better the way in which the
theory cuts itself off.\\
The study of mixed processes, in which a current scaling limit and a hadron
scaling limit are taken at the same time, is being undertaken by several
theorists, including Bjorken.$^{{\rm 29}}$. Such studies may lead us
to a guess as to the general systematics of mixed scaling, that would
include current scaling and hadronic scaling as special cases. That
might well be useful for understanding the connection, if any, between
the two.\\
In the course of such work further attention will no doubt be paid to
the hypothesis of scaling in the hadronic production of lepton pairs; that
is an example of a conjecture about mixed scaling, since the $s$ value of
the initial hadron system and the $q^2$ of the lepton pair are both
supposed to go to infinity, and in proportion.\\
In conclusion, let us express our hope that this summary of problems and
difficulties may encourage some theoretical research and perhaps some
experimental work that will reduce the number of mysteries facing us and
allow the beauty and simplicity of the merging picture of hadrons to stand
out more clearly.\\
\newpage
\noindent
{\bf ACKNOWLEDGEMENTS}\\
\\
We would like to thank J. D. Bjorken, R. P. Feynman, and C. H. Llewellyn
Smith for stimulating conversations about the relation of our work to
previous work on ``partons''. One of us (H.F.) would like to express his
gratitude to the DAAD, to SLAC, and to the AEC high energy physics group
at Caltech for support.\\
The ninth Section, prepared for the Tel Aviv Conference, contains a number
of points that have been elaborated between the Conference and the time
of publication, especially matters concerned with ``anomalies''. For
many enlightening discussions of these questions, we are deeply indebted
to W. Bardeen and to the staff of the Theoretical Study Division of CERN.\\
\\
{\bf APPENDIX K. SCALING HYPOTHESIS}\\
{\bf FOR THE ENERGY MOMENTUM TENSOR}\\
\\
The underlying physical process is the interaction of an off--shell
graviton with a hadronic target, with no momentum transfer and an average
taken over spins. The corresponding matrix element is:
\begin{displaymath}
W_{\mu \nu \delta \sigma} \equiv \frac{1}{4 \pi} \int e^{-iq \cdot z}
< p \mid \left[ \overline{\Theta_{\mu \nu} (z), \Theta_{\rho \sigma} (0)}
 \right] \mid p > d^4 z \, .
\end{displaymath}
\\
We can describe the process by five structure functions.\\
\begin{eqnarray}
W_{\mu \nu \rho \sigma} \left( q^2, q \cdot p \right) & = & \left(
\delta_{\mu \sigma} - \frac{q_{\mu} q_{\nu}}{q^2}  \right)
\left( \delta_{\rho \sigma} - \frac{q_{\rho} q_{\sigma}}{q^2} \right)
T_1 \left( q^2, q \cdot p \right) \nonumber \\ \nonumber \\
& & + \left(  \delta_{(\mu (\rho} - \frac{q_{(\mu} q_{(\rho}}{q^2} \right)
\left( \delta_{\nu) \sigma)} - \frac{q_{\nu)} q_{\sigma)}}{q^2} \right)
T_2 \left( q^2, q \cdot p \right) \nonumber \\ \nonumber \\
& & + P_{\mu} P_{\nu} P_{\rho} P_{\sigma} T_3 \left( q^2, q
\cdot p \right) \nonumber \\ \nonumber\\
& + & \left\{ P_{\mu} P_{\nu} \left( \delta_{\rho \sigma} -
\frac{q_{\rho} q_{\sigma}}{q^2} \right) + P_{\rho} P_{\sigma}
\left( \delta_{\mu \nu} - \frac{q_{\mu} q_{\nu}}{q^2} \right) \right\}
T_4 \left( q^2, q \cdot p \right) \nonumber \\ \nonumber \\
& & + P_{(\mu } P_{(\rho} \left( \delta_{\nu) \sigma)} -
\frac{q_{\nu)}q_{\sigma)}}{q^2} \right) T_5 \left( q^2, q \cdot p \right.
\nonumber
\end{eqnarray}
where
\begin{displaymath}
P_{\mu} \equiv \frac{1}{\sqrt{- q \cdot p}} \left( p_{\mu} -
\frac{q \cdot p}{q^2} q_{\mu} \right) \, .
\end{displaymath}
\end{enumerate}
The symbol () means symmetrization.\\
According to the scaling hypothesis and our principle of higher
dimensions of the symmetry breaking terms, we expect:
\begin{enumerate}
\item[l)] At least one of the dimensionless structure functions
$T_i \left( q^2, q \cdot p \right)$ behaves in the deep inelastic region
like
\begin{displaymath}
T_i \left( q^2, q \cdot p \right) \rightarrow G_i (\xi) \, .
\end{displaymath}
\begin{displaymath}
\zeta = - \frac{q^2}{2 qp}
\end{displaymath}
\item[2)] No structure function diverges in the deep inelastic region.
This is a specific consequence of our postulate about symmetry breaking
effects. Note that (2) is not true in certain Lagrangian models, e. g.,
in a theory with a formal interaction term of dimension-6 like
$\bar\psi \psi \bar\psi \psi$.
\item[3)] The trace terms
\begin{displaymath}
\delta_{\mu \nu} W_{\mu \nu \rho \sigma}, \delta_{\rho \sigma}
W_{\mu \nu \rho \sigma}, \, \, \, {\rm and} \, \, \,
\delta_{\mu \nu} \delta_{\rho \sigma} W_{\mu \nu p \sigma}
\end{displaymath}
are connected with the trace of the energy momentum tensor. The
corresponding structure functions, which can be calculated in terms of
the five functions $T_i$, have to vanish in the deep inelastic region.
That means that we can express the $G_i(\rho )$ in terms of three
non--vanishing structure functions.
\end{enumerate}
\bigskip
\bigskip
{\bf APPENDIX II}\\
\\
We consider deep inelastic current hadron processes in general. Define:
\setcounter{page}{1}
\setcounter{equation}{0}
\renewcommand{\theequation}{A.\arabic{equation}}
\begin{eqnarray}
W_{\mu \nu} \, ^{ij} (q) & = & \frac{1}{4 \pi} \int d^4 z e^{-i q \cdot z}
< p \mid \left[ \overline{{\cal F}_{i \mu} (x), {\cal F}_{j \nu} (y)} 
\right] \mid p > \nonumber \\  \nonumber \\
& = & \left( \delta_{\mu \nu} - \frac{q_{\mu} q_{\nu}}{q^2} \right)
\left( W^{ij}_1 \right. \left( q^2, p \cdot q) \right) -
\frac{\left( q \cdot p \right)^2}{q^2} W^{ij}_2 \left( q^2, p \cdot q
\right) \nonumber \\  \nonumber \\
& & + \frac{\delta_{\mu \nu} \left( p \cdot q \right)^2 +
p_{\mu} p_{\nu} q^2 - \left( p_{\mu} q_{\nu} + p_{\nu} q_{\mu} \right)
\left( p \cdot q \right)}{q^2} W^{ij}_2 \left( q^2, pq \right) \, + \ldots
\nonumber \\
\end{eqnarray}
\\
($p$: arbitrary one--particle state; $z = x - y$).\\
\\
\begin{eqnarray}
W^{5 ij}_{\mu \nu} (q) & = & \frac{1}{4 \pi} \int d^4 z e^{-iq \cdot z}
< p \mid
\left[ \overline{{\cal F}_{i\mu} \, ^5 (x), {\cal F}_{j \nu}
(y)} \right] \mid p > \nonumber \\  \\
& = & - \frac{i}{2} \varepsilon_{\mu \nu \alpha \beta} p_{\alpha}
q_{\beta} W^{5ij}_{3} \left( q^2, p \cdot q \right) + \cdots  \nonumber
\end{eqnarray}
\\
where $\cdots $ denotes terms which destroy the conservation. In the
Bjorken limit, we obtain:\\
\begin{eqnarray}
\lim_{bj} W^{ij}_1 & = & F^{ij}_1 (\xi), \qquad
\lim_{bj} (- p \cdot q) W^{ij}_2 = F_2 \,^{ij} (\xi) \, ,
\nonumber \\ \nonumber \\
\lim_{bj} (- p \cdot q) W^{5ij}_{3}
\left( q^2, q \cdot p \right) & = & F_3^{5ij} (\xi) \, . \nonumber
\end{eqnarray}
\\
We assume $\sigma_L \rightarrow 0$ and get:
\begin{eqnarray}
W^{ij}_{\mu \nu} & \rightarrow &
\frac{\left( p_{\mu} q_{\nu} + p_{\nu} q_{\mu} \right) \left( p \cdot q
\right) - \delta_{\mu \nu} \left( p \cdot q \right)^2 - p_{\mu} p_{\nu}
q^2}{q^2 (q \cdot p)} F_2^{ij} (\xi) \nonumber \\ \nonumber \\
& \rightarrow & \frac
{\left(p_{\mu} q_{\nu} + p_{\nu} q_{\mu} \right)
- \delta_{\mu \nu} (p \cdot q) + 2 p_{\mu}p_{\nu} \xi}{q^2} \,
F_2 \, ^{ij} (\xi) \nonumber \\  \nonumber \\
& \rightarrow & - \frac{s_{\mu \nu \rho z} P_{\sigma} \left( q_{\rho}
+ \xi p_{\rho} \right)}{2 ( q \cdot p)} \frac{F_2^{ij} (\xi)}{\xi} \, .
\end{eqnarray}
\\
Similarly, we find
\begin{equation}
W^{5ij}_{\mu \nu} \rightarrow \frac{1}{2 (q \cdot p)}
\varepsilon_{\mu \nu \sigma \beta} p_{\alpha} q_{\beta} F_3 \, ^{5ij}
(\xi) \, .
\end{equation}
\\
We use the formula of Section V in order to relate $F^{ij}_3$ and
$F_3^{5ij}$ to the bilocal operators appearing there. We find:
\begin{displaymath}
W_{\mu \nu} \, ^{ij} (q) \rightarrow
\frac{s_{\mu \nu \rho \sigma}}{16 \pi^2} \int e^{iq \cdot z} d^4 z
\partial_{\rho} \left( \varepsilon \left( z_0 \right) \delta \left( z^2
\right) \right)
\end{displaymath}
\begin{displaymath}
< p \mid if_{ijk} \left( {\cal F}_{k \sigma} (x, y) + {\cal F}_{k \sigma}
(y, x) \right) + d_{ijk} \left( {\cal F}_{k \sigma} ( x, y) -
{\cal F}_{k \sigma} (y, x) \right) \mid p >
\end{displaymath}
\\
and
\\
\begin{eqnarray}
W_{\mu \nu} \, ^{5i,j} (q) & \rightarrow &
\frac{i \varepsilon_{\mu \nu \rho \sigma}}{16 \pi^3}
\int e^{-iq \cdot z} d^4 z \partial_{\rho} \left( \varepsilon
\left( z_0 \right) \delta \left( z^2 \right) \right) \nonumber \\
\\
& & < p \mid i f_{ijk} \left( {\cal F}_{k \sigma} (y, x) -
{\cal F}_{k \sigma} (x, y) \right) - d_{ijk} \left( {\cal F}_{k \sigma}
(x, y) + {\cal F}_{k \sigma} (y, x) \right) \mid p> \, . \nonumber
\end{eqnarray}
\\
We define:
\begin{eqnarray}
< p \mid {\cal F}_{k \rho} (y, x) - {\cal F}_{k \rho} (x, y) \mid p >
& \equiv & + 2 \tilde{A}^k (p \cdot z) p_{\rho} + \, \, \, {\rm trace \, \, \,
terms}, \nonumber \\  \\
< p \mid {\cal F}_{k \rho} (y, x) + {\cal F}_{k \rho} (x, y) \mid p >
& \equiv & + 2 \tilde{S}^k (p \cdot z) p_{\rho} + \, \, \,
{\rm trace \, \, \, terms} \nonumber
\end{eqnarray}
\\
where $z = x - y$ is light--like,
\\
\begin{eqnarray}
\tilde{A}^k (p, z) & \equiv & \int e^{-i \xi (p \cdot z)} A^k (\xi) d \xi,
\nonumber \\  \nonumber \\
\tilde{S}^k (p \cdot z) & \equiv & \int e^{-i \xi (p \cdot z)}
S^k (\xi) d \xi \, .
\end{eqnarray}
\\
Inserting (A.9) into (A.8) and (A.7), we obtain
\begin{eqnarray}
W^{ij}_{\mu \nu} (q) & \rightarrow & +
\frac{s_{\mu \nu \rho \sigma} p_{\sigma}}{8 \pi^2} \int d^4 z
e^{- i q \cdot z} \partial_{\rho} \left( \varepsilon \left( z_0 \right)
\delta \left( z^2 \right) \right) \nonumber \\
& & \left\{ if_{ijk} \tilde{S}^k (p \cdot z) + d_{ijk} \tilde{A}^k
(p \cdot z) \right\} \, , \nonumber \\ \nonumber \\
W^{5i, j}_{\mu \nu} (q) \rightarrow & + &
\frac{i \varepsilon_{\mu \nu \rho \sigma}}{8 \pi^2} \int d^4 z
e^{- i q \cdot z}
\partial_{\rho} \left( \varepsilon \left( z_0 \right) \delta \left(
z^2 \right) \right) \nonumber \\ \nonumber \\
& & \left\{ if_{ijk} \tilde{A}^k (p \cdot z) - d_{ijk} \tilde{S}^k
(p \cdot z) \right\} \, , \nonumber
\end{eqnarray}
\\
and get further, using (A.9),
\begin{eqnarray}
W_{\mu \nu} \, ^{ij} \rightarrow & + &
\frac{s_{\mu \nu \rho \sigma} p \sigma}{8 \pi^2} \int d \alpha \int d^4 z
\, e^{-i (q + \alpha \cdot p)
\cdot z} \partial_{\rho} \left( \varepsilon \left( z_0 \right) \delta
\left( z^2 \right)  \right) \nonumber \\  \nonumber \\
& & \left\{ if_{ijk} S^k (a) + d_{ijk} A^k (a) \right\} \nonumber \\
\nonumber \\
& = & - \frac{i s_{\mu \nu \rho\sigma} p \sigma}{8 \pi^2} \int
d \alpha \int d^4 z (q + \alpha \cdot p)_{\rho} \,
e^{-i (q + \alpha \cdot p) \cdot z}
\, \varepsilon \left( z_0 \right) \delta \left( z^2 \right) \nonumber \\
\nonumber \\
& & \left\{ i f_{ijk} S^k (a) + d_{ijk} A^k (a) \right\} \nonumber \\
\\
& = & - \frac{s_{\mu \nu \rho \sigma} p_{\sigma}
\left( q_{\rho} + \xi p_{\rho} \right)}{4 ( q \cdot p)}
\left\{ if_{ljk} S^k (\xi) + d_{ljk} A^k (\xi) \right\} \, , 
\nonumber \\ \nonumber \\
\xi & = & - \frac{q^2}{2 (q \cdot p)} \, .  \nonumber
\end{eqnarray}
\\
Similarly, we obtain
\begin{equation}
W^{5i, j}_{\mu \nu} (q) \rightarrow +
\frac{ i \varepsilon_{\mu \nu \rho \sigma}
p_{\rho} q_{\sigma}}{4 (q \cdot p)}
\left\{ i f_{ijk} A^k (\xi) - d_{ijk} S^k (\xi)  \right\} \, .
\end{equation}
\\
We compare (A.12) with (A.6) and find
\begin{equation}
F^{ij}_2 (\xi) = + \frac{1}{2} \xi \left\{ if_{ijk} S^k (\xi) + d_{ijk}
A^k (\xi) \right\}
\end{equation}
\\
and also (13) with (7)
\begin{equation}
F_3 \, ^{5ij} (\xi) = + \frac{1}{2} \left\{ if_{ijk} A^k (\xi) -
d_{ijk} S^k (\xi) \right\} \, .
\end{equation}
\\
Further, we should like to demonstrate how can one compute the functions
$S^k (\xi)$ and $A^k (\xi)$ in terms of the expectation values of the
local operators appearing in the Taylor expansion of the bilocal
operators.\\
Using the definitions (A.8) and neglecting internal indices, we get:
\begin{eqnarray}
{\cal F} \, _{\rho} (y, x) + {\cal F} \, _{\rho} (x, y) & \sim &
i \bar q (y) \gamma_{\rho} q (x) + i \bar q (x) \gamma_{\rho} q (y)
\nonumber \\  \nonumber \\
& \sim & 2 i \bar q (x) \gamma_{\rho} q (x) - z^{\alpha} i \left( \bar q
(x) \gamma_{\rho} \partial_{\alpha} q (x) + \partial_{\alpha}
\bar q (x) \gamma_{\rho}(x) \right) \nonumber \\  \nonumber \\
& + & \frac{i}{2!} z^{\alpha} z^{\beta} \left( \bar q (x) \gamma_{\rho}
\partial_{\alpha} \partial_{\beta} q (x) + \partial_{\alpha} 
\partial_{\beta} \bar q (x) \gamma_{\rho} q (x) \right) + \cdots
\nonumber
\end{eqnarray}
\\
We define the numbers $s_i$:
\begin{displaymath}
< p \mid ''i \bar q (x) \gamma_{\rho} q (x) '' \mid p > = s_1 p_{\rho}
\end{displaymath}
\begin{displaymath}
\frac{1}{2} < p \mid ''\bar q (x) \gamma _{\rho} \partial_{\alpha}
\partial_{\beta} q (x) + \partial_{\alpha} \partial_{\beta} \bar q (x)
\gamma_{\rho} q (x)'' \mid p >
\end{displaymath}
\begin{equation}
= s_3 p_{\rho} p_{\alpha} p_{\beta} + \, \, {\rm trace} \, \, {\rm terms,}
\, \, etc.
\end{equation}
\\
and find\\
\begin{equation}
\tilde{S}^k (p \cdot z) = + \left( s_1 \, ^k + \frac{s_3^k}{2!}
(p \cdot z)^2 + \cdots \right)
\end{equation}
\\
Similarly we define the dimensionless number $\alpha_i$:
\begin{displaymath}
\frac{1}{2} < p \mid '' \bar{q} (x) \gamma_{\rho} \partial_{\alpha} q (x)
- \partial_{\alpha} \bar{q} (x) \gamma_{\rho} q(x)'' \quad \mid p > =
a_2 p_{\rho} p_{\alpha} \cdots
\end{displaymath}
\\
and find\\
\begin{equation}
\tilde{A}^k (p \cdot z) = + i \left( a_2^k ( p \cdot z) + \frac{1}{3!}
a_4^k (p \cdot z)^3 + \cdots \right) \, .
\end{equation}
\\
If we carry out the Fourier transform, we obtain, restoring $SU_3$
indices,
\begin{eqnarray}
S^k (\xi) & = & s_1 \, ^k (\xi) - \frac{1}{2!} s_3 \, ^k \delta '' (\xi)
+ \cdots , \\  \nonumber \\
A^k (\xi) & = & \left( a_2 \, ^k \delta'' (\xi) -
\frac{1}{3!} a_4 \, ^k \delta''' (\xi) + \ldots \right) \, .
\end{eqnarray}
\\
{\bf REFERENCES}\\
\begin{enumerate}
\item[1.] K. G. WILSON, {\it Phys. Rev.} {\bf 179}, 1499 (1969).
\item[2.] R. BRANDT AND G. PREPARATA, CERN preprint TH--1208.
\item[3.] Y. FRISHMAN, Phys. Rev. Lett. {\bf 25}, 966 (1970).
\item[4.] R. P. FEYNMAN, {\it Proceedings of Third High Energy Collision
Conference at State University of New York}, Stony Brook, Gordon and
Breach, 1970.
\item[5.] J. D. BJORKEN AND E. A. PASCHOS, Phys. Rev. {\bf 185}, 1975
(1969).
\item[6.] P. LANDSHOFF AND J. C. POLKINGHORNE, Cambridge University DAMTP
preprints (1970).
\item[7.] C. H. LLEWELLYN SMITH, Nucl. Phys. {\bf B17}, 277 (1970).
\item[8.] S. D. DRELL AND T. YAN, Phys. Rev. Lett. {\bf 24}, 181 (1970).
\item[9.] Note we use the metric $\delta_{\mu \nu} = (1, 1, 1, -1)$ and
the covariant state normalization $< p' s' \mid p s > = (2 \pi)^3 2 p_0
\delta \left( p - p' \right) \delta_{s's}$.
\item[10.] M. GELL--MANN, {\it Proceedings of Third Hawaii Topical
Conference on Particle Physics}, Western Periodicals Co., Los Angeles,
1969.
\item[11.] F. VON HIPPEL AND J. K. KIM, {\it Phys. Rev.} {\bf D1}, 151
(1970).
\item[12.] J. ELLIS, Physics Lett. {\bf 33B}, 591 (1970).
\item[13.] R. F. DASHEN AND T. P. CHENG, Institute for Advanced Study
preprint (1970).
\item[14.] J. ELLIS, P. WEISZ, AND B. ZUMINO, {\it Phys. Lett.} {\bf 34B},
91 (1971).
\item[15.] E. D. BLOOM, G. BUSCHORN, R. L. COTTRELL, D. H. COWARD, H.
DESTAEBLER, J. DREES, C. L. JORDAN, G. MILLER, L. Mo, H. PIEL, R. E.
TAYLOR, M. BREIDENBACH, W. R. DITZLER, J. I. FRIEDMAN, G. C. HARTMANN,
H. W. KENDALL, AND J. S. POUCHER, Stanford Linear Accelerator Center
preprint SLAC--PUB--796 (1970) (report presented at the XVth International
Conference on High Energy Physics, Kiev, USSR, 1970).
\item[16.] H. LEUTWYLER AND J. STERN, {\it Nucl. Phys.} {\bf B20}, 77
(1970); R. JACKIW, R. VAN ROYEN, AND G. B. WEST, {\it Phys. Rev.}
{\bf D2}, 2473 (1970).
\item[17.] S. CICCARIELLO, R. GATTO, G. SARTORI, AND M. TONIN, {\it Phys.
Lett.} {\bf 30B}, 546 (1969); G. MACK, {\it Phys. Rev. Lett.} {\bf 25},
400 (1970). J. M. CORNWALL AND R. E. NORTON, {\it Phys. Rev.} {\bf 177},
2584 (1968) used a different approach to accomplish about the same result.
Instead of expanding light cone commutators, they use equal time
commutators with higher and higher time derivatives, sandwiched between
states at infinite momentum. That amounts to roughly the same thing, and
represents an alternative approach to light cone algebra.
\item[18.] C. H. LLEWELLYN SMITH, to be published.
\item[19.] R. P. FEYNMAN, M. GELL--MANN, AND G. ZWEIG, {\it Phys. Rev.
Lett.} {\bf 13}, 678 (1964). The idea was applied to many important
effects by J. D. BJORKEN, Phys. Rev. {\bf 148}, 1467 (1966).
\item[20.] J. MANDULA, A. SCHWIMMER, J. WEYERS, AND G. ZWEIG have proposed
independently this test of the dimension $l_{\mu}$ and are publishing a
full account of it.
\item[21.] S. ADLER, {\it Phys. Rev.} {\bf 143}, 154 (1966).
\item[22.] D. J. GROSS AND C. H. LLEWELLYN SMITH, {\it Nucl. Phys.}
{\bf B14}, 337 (1969).
\item[23.] H. SUGAWARA, {\it Phys. Rev.} {\bf 170}, 1659 (1968).
\item[24.] J. M. CORNWALL AND R. JACKIW, UCLA preprint (1971).
\item[25.] D. J. GROSS AND S. B. TREIMAN, Princeton University preprint
(1971).
\item[26.] W. BARDEEN, private communication.
\item[27.] M. GELL--MANN AND F. E. LOW, {\it Phys. Rev.} {\bf 95}, 1300
(1954); M. BAKER AND K. JONSON, {\it Phys. Rev.} {\bf 183}, 1292 (1969).
\item[28.] A. H. MULLER, {\it Phys. Rev.} {\bf D2}, 2963 (1970).
\item[29.] J. D. BJORKEN, Talk given at the same Conference.
\end{enumerate}
\end{document}